\pdfoutput=1
\documentclass[journal]{IEEEtran}

\ifCLASSINFOpdf
  \usepackage[pdftex]{graphicx}
\else
  \usepackage[dvips]{graphicx}
\fi

\usepackage{amsmath,amssymb,amsfonts}

\usepackage{array}

\usepackage[stretch=20,shrink=40]{microtype}

\usepackage{xurl}
\usepackage[hidelinks]{hyperref}
\usepackage{placeins}
\usepackage{dblfloatfix}

\usepackage{afterpage}
\usepackage{algpseudocode}
\usepackage{algorithm}
\usepackage{textcomp}
\usepackage{cleveref}
\crefname{section}{Section}{Sections}
\crefname{subsection}{Section}{Sections}
\crefname{subsubsection}{Section}{Sections}
\crefname{equation}{Eq.}{Eqs.}
\crefname{lemma}{Lemma}{Lemmas}
\Crefname{lemma}{Lemma}{Lemmas}
\usepackage{float}
\usepackage{bm}
\usepackage{bbold}

\usepackage{amsthm}

\usepackage{threeparttable}

\newtheorem{lemma}{Lemma}

\usepackage[table,dvipsnames]{xcolor}

\usepackage{booktabs}
\usepackage{makecell}
\usepackage{multirow}

\newcommand{\stkcell}[2]{\shortstack{#1\\[-2pt]{\scriptsize\color{black!65}#2}}}
\newcommand{\stkbest}[2]{\cellcolor{green!22}\shortstack{\textbf{#1}\\[-2pt]{\scriptsize\color{black!65}#2}}}
\newcommand{\stksec}[2]{\cellcolor{green!12}\shortstack{\underline{#1}\\[-2pt]{\scriptsize\color{black!65}#2}}}
\newcommand{\stkthird}[2]{\cellcolor{green!5}\shortstack{#1\\[-2pt]{\scriptsize\color{black!65}#2}}}

\usepackage{xr}

\begin{document}
\AddToHookNext{shipout/foreground}{\put(0,0){\makebox[0pt][l]{\raisebox{-\dimexpr\paperheight-22pt\relax}{\makebox[\paperwidth]{\parbox{\textwidth}{\centering\scriptsize This work has been submitted to the IEEE for possible publication. Copyright may be transferred without notice, after which this version may no longer be accessible.}}}}}}
\title{SpiderSkel: Sub-Pixel Multi-Tangent 2D Skeletonization from Inscribed-Disc Geometry}
\hypersetup{pdftitle={SpiderSkel: Sub-Pixel Multi-Tangent 2D Skeletonization from Inscribed-Disc Geometry},
  pdfauthor={Zhexian Zhou, Lian Ren, Xiao Feng, Sebastian Scherer}}

\author{Zhexian~Zhou,
        Lian~Ren,
        Xiao~Feng,
        and~Sebastian~Scherer,~\IEEEmembership{Senior~Member,~IEEE}%
\thanks{Corresponding authors: Zhexian Zhou and Sebastian Scherer.}%
\thanks{Z. Zhou is with the Robotics Institute, Carnegie Mellon University,
Pittsburgh, PA 15213 USA, and was previously with the Department of
Electrical and Computer Engineering, Carnegie Mellon University,
Pittsburgh, PA 15213 USA
(e-mail: jakozhou@gmail.com; ORCID: 0009-0004-8278-1230).}%
\thanks{L. Ren was with the Department of Electrical and Computer Engineering,
Carnegie Mellon University, Pittsburgh, PA 15213 USA
(e-mail: renglian@outlook.com; ORCID: 0009-0006-6512-572X).}%
\thanks{X. Feng is with the National Key Laboratory of Wireless Communications,
University of Electronic Science and Technology of China, Chengdu, Sichuan
610054, China
(e-mail: fengxiaocd@gmail.com; ORCID: 0000-0002-8857-9760).}%
\thanks{S. Scherer is with the Robotics Institute,
Carnegie Mellon University, Pittsburgh, PA 15213 USA
(e-mail: basti@andrew.cmu.edu; ORCID: 0000-0002-8373-4688).}}

\markboth{}{}

\maketitle
\bstctlcite{IEEEtran:BSTcontrol}

\begin{abstract}
Skeletonization reduces a binary shape to a compact medial representation,
but pixel-grid discretization perturbs the boundary-contact geometry
underlying the medial axis, creating tension among connectivity,
reconstruction fidelity, and skeleton compactness. SpiderSkel is a family of
training-free two-dimensional skeletonization methods built on multi-tangent
inscribed-disc geometry. A tie-aware exact Euclidean distance transform
retains tied nearest boundary sites rather than collapsing them to a single
label; medial candidates are localised to sub-pixel positions from the
inscribed disc's tangent contacts, filtered by the angular span of those
contacts, and completed by connectivity-aware refinement. Across nineteen
public benchmarks comprising 11,323 images of shapes, retinal vessels, cracks
and roads, and document images, a single operating point fixed across
datasets yields skeletons that, on every image, have exact skeleton topology under
the component/hole certificate, remain fully connected, and have exact
per-object reconstruction topology. The three SpiderSkel variants are the only evaluated methods to
satisfy all three properties across every category; within each category,
every variant is Pareto-optimal with respect to all twenty-two baselines in
strict sub-pixel reconstruction Dice versus reduction rate. Incremental
engines for all three variants reproduce the skeleton of a complete
recomputation exactly across 23,040 verified edit steps, with median update
latencies of
15.4--18.1 ms per foreground edit and 4.3--4.4 ms per canvas edit, and speedups
reaching 33.0-fold on canvas edits.
\end{abstract}

\begin{IEEEkeywords}
Digital topology, distance transform, medial axis, skeletonization, sub-pixel
interpolation
\end{IEEEkeywords}

\IEEEpeerreviewmaketitle

\begin{figure}[!t]
  \centering
  \includegraphics[width=\linewidth]{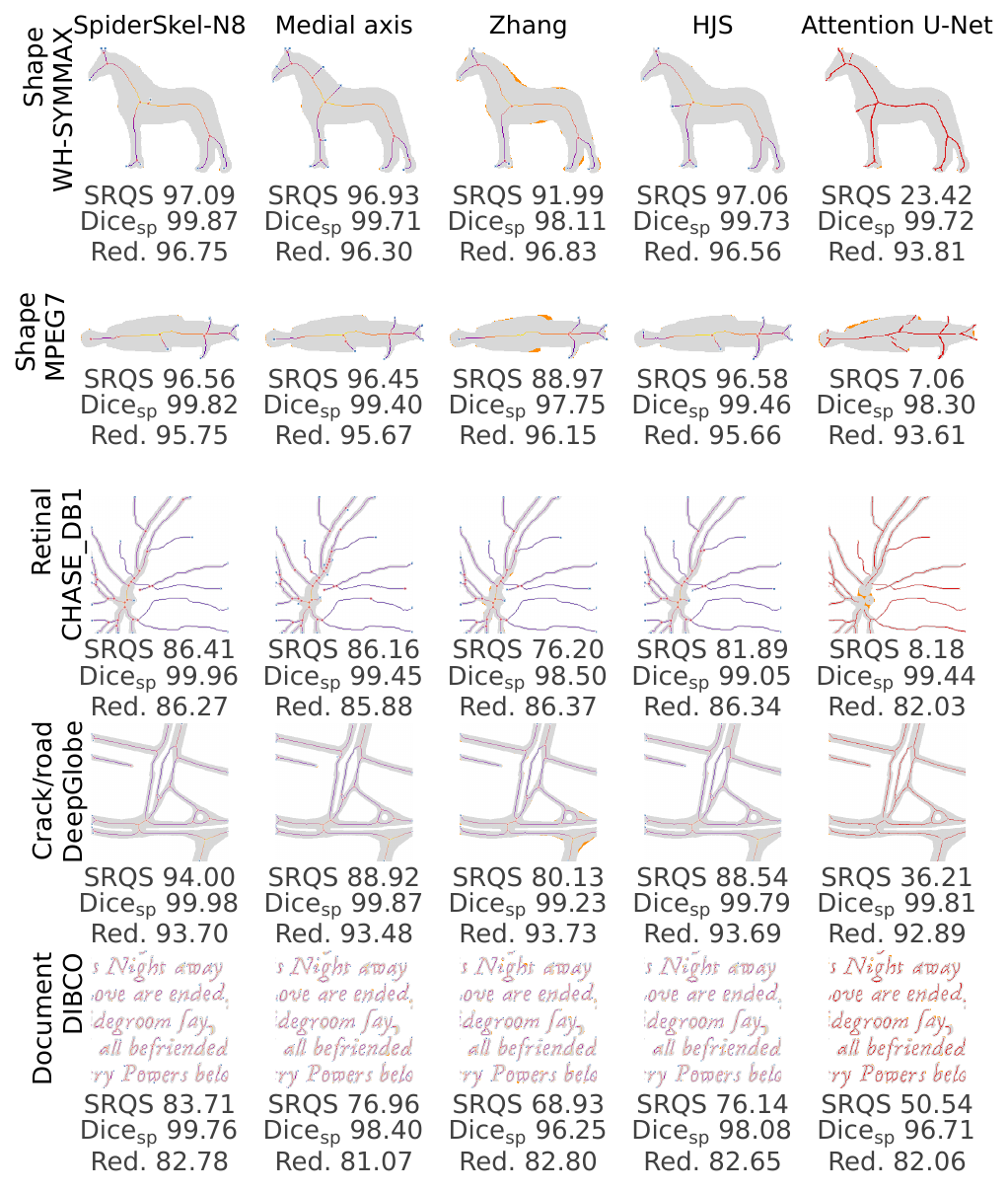}
  \caption{Diagnostic qualitative comparison across the four benchmark
  categories. Columns: SpiderSkel-N8 and one representative
  each of the medial-axis, thinning, flux-based and learned families; the
  gray silhouette in every cell is the input mask. Geometric skeletons are coloured by local radius, from thick
  to thin, with junctions in red and endpoints in blue; the Attention U-Net
  column shows the raw network output, without thinning post-processing.
  Orange marks mask pixels the reconstruction misses. Scores below each
  cell are computed on the rendered cell: SkelReconQS (printed SRQS),
  strict sub-pixel reconstruction Dice (Dice$_{\mathrm{sp}}$) and Reduction
  Rate (Reduct.), all \%, defined in Section~\ref{subsubsec:srqs_metric}.
  Thin-structure rows are shown at higher magnification. All rows are held
  out from the learned baseline's training data and are hand-picked for
  illustration; Table~\ref{tab:category_avg_skelqs_em} gives the
  quantitative comparison.}
  \label{fig:method_matrix}
\end{figure}

\section{Introduction}

\IEEEPARstart{S}{keletonization} compresses an object to a minimal medial representation
while aiming to preserve its topology and geometry, a problem studied since
Blum introduced the medial axis in 1967~\cite{blum1967MAT}. It underpins applications in
biological image analysis \cite{DBLP:journals/pami/DillLN87}, terrain reconstruction \cite{skeleton_terrain}, animation \cite{skeleton_animation}, 3D model recognition \cite{3D_skeleton_correct}, and 3D reconstruction \cite{lee2013skeleton}.

The central difficulty is that the skeleton is a global geometric object, yet
it must be recovered from a discrete grid that perturbs tangent locations.
Classical grid-based skeletonization follows two dominant strategies,
iteratively peeling the object from its contour under template or topological
tests \cite{thinning_survey, arcelli_paral_thinning, CHU1986207} and extracting
local maxima of a distance transform \cite{seq_op, arcelli_from_local_maximas};
both must summarise global geometric information through local
computation.

Exact Euclidean distance transforms (EDTs) \cite{meijster2000general}, exact
multi-site feature-transform sets \cite{DBLP:journals/ipl/Hesselink07} and
single-representative closest-point skeletons \cite{luo2011euclidean,
DBLP:journals/pami/HesselinkR08} are established prior art. The gap we
address lies downstream: how the retained, possibly multi-valued
nearest-site information is used. Section~\ref{sec:Related Work} positions this against the classical, continuous-geometry, nearest feature-transform and closest-point families, and Fig.~\ref{fig:method_matrix} compares representative methods
from each family on the same masks.

We therefore ask whether an exact EDT-based skeletonization can use the
multi-valued nearest-site field to mitigate discretization-induced
misalignment and produce a stable sub-pixel medial representation with
controlled connectivity, reconstruction and reduction. We answer with our
training-free method family,
\emph{SpiderSkel}, in three variants (SpiderSkel-N8, SpiderSkel-DualC and
SpiderSkel-Scaffold).

\noindent\textbf{Our contributions are:}
(i)~a tie-aware augmentation of \textit{Meijster}'s exact
EDT~\cite{meijster2000general} that enumerates all tied nearest boundary
sites and retains a deterministic, bounded set of representatives for
downstream interpolation, with sensitivity to the cap reported;
(ii)~a \emph{subpixel multi-tangent medial interpolation}, grounded in Blum's
medial axis~\cite{blum1967MAT} and the bisector geometry of Voronoi
partitions~\cite{voronoi1908nouvelles}, that localises sub-grid medial candidates at
the perpendicular-bisector intersections of paired tangents across a
geometry-adaptive neighbourhood;
(iii)~a tangent-angle-span filter and support-point sampling test that
provide continuous significance control in the classical object-angle
tradition~\cite{DBLP:journals/ivc/MalandainF98, DBLP:conf/sma/FoskeyLM03,
DBLP:journals/ivc/CouprieCZ07}, together with explicitly specified
connection, cleaning, thinning and optional reduction stages;
(iv)~incremental engines for all three variants, verified to reproduce
the skeleton of a complete recomputation exactly over $23{,}040$ recorded
edit steps; and
(v)~a nineteen-benchmark evaluation at one fixed operating point, in which
the three variants are the only evaluated methods attaining exact skeleton
topology (per-object agreement in connected-component and hole counts),
full connectivity and exact per-object reconstruction topology in every
category aggregate, and each is Pareto-optimal against all baselines in the
trade-off between strict sub-pixel reconstruction Dice and Reduction Rate in
all four categories. Under a twelve-condition robustness protocol, they further achieve the
highest pooled absolute score among the nine evaluated methods in every
condition, with no per-condition mean change from the clean-input score below
$-1.9$~pp.

SpiderSkel-Scaffold additionally preserves each object's component and cycle
counts by construction (Section~\ref{subsec: Algorithm Pipeline}).
 
\section{Related Work}\label{sec:Related Work}

\subsection{Distance Transform and Exact Feature Sets}

The distance transform (DT), which assigns each pixel its smallest distance to the background, was introduced by \textit{Rosenfeld and Pfaltz} \cite{seq_op}; later work traded exactness for cost through
quasi-Euclidean metrics \cite{Borgefors_quasi_EDT, dan_pseudo_EDT,
DT_arbitary_dimension}; for distance-based skeletonization, however, the true
Euclidean metric avoids the anisotropy such approximations introduce, and
pseudo-EDTs propagating through a single neighbourhood remain inexact
\cite{DT_2D_survey}. Exact EDT algorithms
followed \cite{SAITO_EDT, meijster2000general, C&M_EDT, Maurer_EDT,
L&Z_EDT_morph}; \textit{Meijster} \cite{meijster2000general} refines
\cite{SAITO_EDT} into a linear-time exact EDT and is the basis for our envelope
construction.

An exact distance value does not identify a unique nearest boundary site at a
Voronoi tie. \textit{Hesselink}'s feature-transform-set algorithm computes the
full set of nearest-site minimisers in linear time and arbitrary dimension
\cite{DBLP:journals/ipl/Hesselink07}. \textit{Hesselink and Roerdink} show that
one lexicographically selected exact feature point per grid site suffices for
their linear-time integer-medial-axis construction, and explicitly do not claim
topology preservation \cite{DBLP:journals/pami/HesselinkR08}. \textit{Luo,
Guibas and Zhao} propagate one approximately closest point per grid site by
fast sweeping and combine the result with distance-ordered homotopic thinning
\cite{luo2011euclidean}. \textit{Telea and van Wijk}'s augmented fast-marching
method likewise propagates a collapsed boundary parameterisation and prunes by
boundary-length significance~\cite{DBLP:conf/vissym/TeleaW02}.

We therefore treat both exact multi-site feature transforms and closest-point
skeletonization as established prior art, and locate our contribution downstream.
Table~\ref{tab:prior_art_claims} isolates the remaining differences
claim by claim. Although the construction generalises beyond two dimensions,
this paper develops and evaluates it empirically in 2D.

\begin{table*}[!t]
\centering
\scriptsize
\setlength{\tabcolsep}{2.5pt}
\renewcommand{\arraystretch}{1.0}
\caption{Claim-by-claim scope comparison with the nearest exact feature-transform and closest-point skeleton literature ($N$ pixels, $N_f$ foreground, $N_{\max}$ cap). Exact feature transforms are prior work; ours differs in the steps that follow}
\label{tab:prior_art_claims}
\begin{tabular}{@{}>{\raggedright\arraybackslash}p{0.115\textwidth}>{\raggedright\arraybackslash}p{0.205\textwidth}>{\raggedright\arraybackslash}p{0.205\textwidth}>{\raggedright\arraybackslash}p{0.205\textwidth}>{\columncolor{black!8}}>{\raggedright\arraybackslash}p{0.205\textwidth}@{}}
\toprule
 & Hesselink (2007)~\cite{DBLP:journals/ipl/Hesselink07} & Hesselink--Roerdink (2008)~\cite{DBLP:journals/pami/HesselinkR08} & Luo--Guibas--Zhao (2011)~\cite{luo2011euclidean} & \textbf{This work} \\
\midrule
Computed output & Exact feature-transform set & Exact IMA skeleton & Closest-point skeleton & Exact distances + tangent set \\
Treatment of ties & All nearest sites & One lexical minimiser & Approximate; multiplicity inferred & All tied nearest sites \\
Complexity & Linear time & Linear time & Iterative fast sweeping & $\mathcal{O}(N{+}T)$; $T$ = tie mass \\
Dimension & Arbitrary & Arbitrary (2D/3D shown) & 2D/3D & 2D study \\
Memory & Linear arrays plus feature lists & Coordinate vector per point & Distance, closest-point grids & $O(N+N_fN_{\max})$ \\
Interpolation & No & No & No & Yes, multi-tangent subpixel \\
Topology reduction & None & IMA pruning; no guarantee & DT-ordered homotopic thinning & Connectivity-aware heuristic \\
\bottomrule
\end{tabular}
\end{table*}
 
\subsection{Skeletonization Algorithms}

Since \textit{Blum}'s medial axis transformation \cite{blum1967MAT}, much work has pursued representative or application-oriented skeletons from 2D digital images \cite{thinning_survey}. The dominant strategy iteratively
``peels'' contour pixels, retaining those that satisfy template-matching
\cite{arcelli_paral_thinning, CHU1986207} or local-maximum
\cite{arcelli_from_local_maximas, DBLP:journals/isci/TsaoF82} criteria, with
deletion governed by curated templates \cite{op_pattern_process} or by
topological predicates such as the crossing number \cite{Rutovitz_PR}, the
connectivity number \cite{hilditch1969linear} or simplicity
\cite{skeleton_1pass2op}. Correct templates and post-processing are themselves hard to design
\cite{3D_skeleton_correct}.
Refinements include connectivity-preserving selection from local maxima, vector-integral methods
that preserve the natural parts hierarchy \cite{untangling_MAT}, and
contour-based branch pruning \cite{DBLP:journals/pami/BaiLL07}; the 3D
literature follows the same outline \cite{DT_3D_paraboloid_medical,
3D_fully_paral_thinning, DBLP:journals/pami/ArcelliBS11}.

Two further classical families extract the skeleton from continuous geometry
rather than iterative peeling, and both are ranked in our benchmark.
Flux-based methods detect medial points where the average outward flux of the
distance-gradient field collapses: \textit{Siddiqi et al.}'s
Hamilton--Jacobi skeletons~\cite{DBLP:journals/ijcv/SiddiqiBTZ02} combine that
criterion with homotopy-preserving thinning, and the family's native
significance parameter is its flux threshold (the $\gamma$ of the reference
implementation~\cite{DBLP:journals/ipol/HeKA21}, whose corpus-level sensitivity
we audit in Supplementary Appendix~\ref{app:hjs_gamma}). Voronoi skeletons instead
prune the Voronoi diagram of boundary samples to a connected medial subgraph
ordered by residual-based significance, as in \textit{Ogniewicz and
K{\"u}bler}'s hierarchic formulation~\cite{DBLP:journals/pr/OgniewiczK95}.

\subsection{Learning-based Skeletonization}

Separate work casts skeletonization as supervised per-pixel labelling. \textit{Shen et al.} \cite{DBLP:journals/tip/ShenZJWBY17} train a multi-task fully convolutional network with scale-associated side outputs to extract object skeletons from natural images. \textit{Ke et al.} \cite{DBLP:conf/cvpr/KeCJZY17} use a side-output residual network with deep supervision, and \textit{Zhao et al.} \cite{DBLP:conf/ijcai/ZhaoSGLC18} add a hierarchical feature-integration module that captures skeletons across scales in a single pass. A complementary paradigm predicts a dual potential (flux) field whose singularities trace the skeleton~\cite{DBLP:conf/cvpr/WangXTBDS19}.

These methods train on, and their accuracy depends on, hand-curated reference skeleton annotations such as Pixel SkelNetOn \cite{DBLP:conf/cvpr/DemirHLMRPFBDK19}. The nineteen benchmarks evaluated here provide only binary object masks, no canonical reference skeleton annotation to supervise against, and quality is assessed through reconstruction rather than skeleton overlap, so we focus on the training-free, geometry-driven regime. On the benchmarks with curated reference skeleton annotations (the shape-matching datasets of \cite{DBLP:journals/ijcv/YangISGKBYG24}), we additionally retrain two learned skeletonizers in-domain: Skelite~\cite{DBLP:conf/ipmi/VargasMPNA25}, an iterative-erosion network, and the attention U-Net of~\cite{DBLP:conf/iccvw/Nguyen21}, a segmentation-style model. We compare them against our method on a held-out split in Supplementary Appendix~\ref{app:learned_comparison}. Even with that supervision the learned models break topology and do not surpass our method on the composite reconstruction-quality score, and they require reference skeleton annotations available for only twelve of the nineteen benchmarks; this two-model comparison on the supervised subset is evidence within that scope, not a universal ranking.
 
\section{Method}\label{sec:Proposed Method}
\subsection{Tangent Geometry of the Exact Distance Transform}
\label{sec:Tangent Points Geometric Basics}
\label{subsec: Discrete Geometry-adaptive Extend Vectors}

The exact EDT assigns each interior point
the radius of the largest inscribed ball centred at it, and the boundary sites realising that
radius are its tangent contacts. Let $\partial\Omega$ denote the discrete
outer boundary ring of the foreground $\Omega$: the background pixels adjacent
to $\Omega$, which the exact EDT measures against. Write
$d(x)=\inf_{Q'\in\partial\Omega}\|x-Q'\|$ for the distance-to-boundary function
and
\begin{equation}\label{eq:tangent_set}
\mathcal{Q}_P = \bigl\{\,Q \in \partial\Omega \ \big|\ \|P-Q\| = d(P)\,\bigr\}
\end{equation}
for the set of nearest boundary sites of an interior point $P$. On a grid this
set is frequently not a singleton, and the method exploits that multiplicity: a point with $|\mathcal{Q}_P|\ge 2$ lies on a discrete
bisector. Every member of $\mathcal{Q}_P$ lies at exactly $d(P)$, so
$\mathcal{Q}_P$ is one distance shell; \cref{subsec: Geometry-adaptive Neighbourhood Feature Aggregation} carries that property to off-lattice points. Quantization also perturbs where the contact transition falls along
the inward normal, so candidates are evaluated jointly over a local
neighbourhood, not from a single interior sample; Supplementary
Appendix~\ref{app:tangent_basics_full} gives the two-phase normal-line
derivation.

We quantify the spread of a site set by an angular span: for a point $x$
and a finite site set $\mathcal{Q}$, sort the polar angles of the sites
around $x$, $\theta_{1}\le\dots\le\theta_{n}$, with $\Delta\theta_{i}$ the
consecutive cyclic gaps, the last wrapping through $2\pi$. The span
operator and its grid-point instance are
\begin{equation}\label{eq:span}
S(x;\mathcal{Q}) = 2\pi - \max_{1\le i\le n} \Delta\theta_{i},
\qquad
S_P = S(P;\mathcal{Q}_P).
\end{equation}

Digital extension from a boundary site toward an interior point is likewise
set-valued: rounding lets several directions reach the same grid point, and
the next rounded sample is not unique on the secondary axes, so the method
enumerates the admissible direction set, not one deterministic ray; Supplementary Appendix~\ref{app:incircle_full} gives the
algebra.

Based on the modelling above, we integrate boundary-derived features into a
primary 2D skeletonizer. Its key mechanism, \emph{subpixel multi-tangent
medial interpolation}, pairs tie-aware tangent candidates from neighbouring
grid cells and intersects their equal-distance locus with the segment
between the cells (\cref{lem:two_site_interp}); there it keeps only the
nearest participating sites, so the span is read on a single distance
shell, and each surviving candidate is qualified by an angular-span filter and
a deterministic support-sampling test. The connectivity filling, cleaning, thinning and
reduction steps that follow are discrete heuristics: connectivity-aware refinements, not a homotopy-preserving transformation of the pipeline. The subsections below define each
construction; \cref{subsec: Algorithm Pipeline} composes them.

\begin{figure*}[!t]
  \centering
  \includegraphics[width=\textwidth]{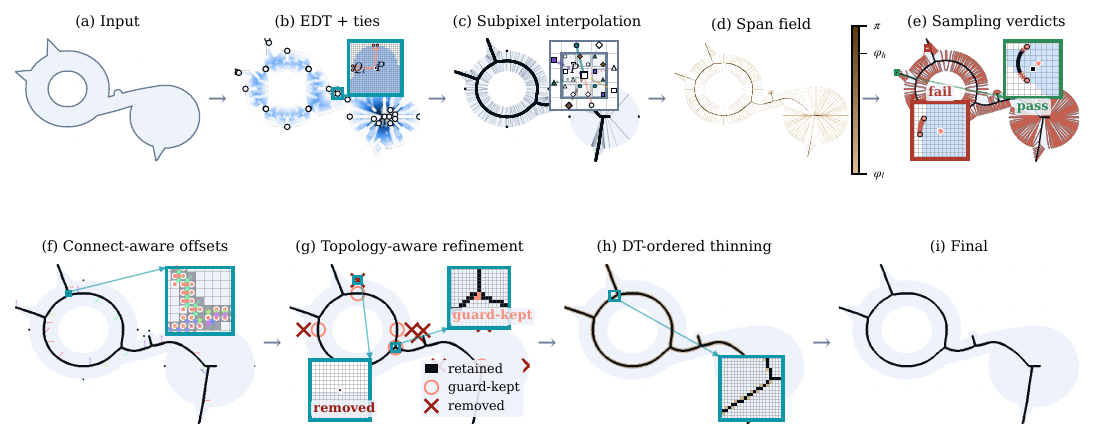}
  \caption{The reported 2D pipeline (\cref{eq:complete_method}) at the default
  operating point $(\phi_l,\phi_h)=(0,\,13\pi/16)$: nine panels trace the
  SpiderSkel-N8 path, input mask to skeleton, one
  execution of the implementation on a constructed
  exemplar not drawn from any benchmark dataset; SpiderSkel-DualC and
  SpiderSkel-Scaffold diverge as \cref{subsec: Algorithm Pipeline} specifies. (a)~Input mask. (b)~Exact EDT,
  multi-valued nearest-site pixels marked, zooming one tied pixel $P$, its
  working set and its geometry-adaptive stencil
  (\cref{subsec: Geometry-adaptive Neighbourhood}). (c)~Subpixel
  interpolation, each kept candidate $\tilde{P}$ at its interpolated
  position with anchor threads to its two tangent sites; inset: the adaptive
  neighbourhood's tangent-line schematic. (d)~The tangent-angle span field, cutoffs $\phi_l,\phi_h$ on
  the colourbar (\cref{eq:phi_thresholds}). (e)~Support-sampling verdicts,
  green (pass) and red (fail) insets replaying
  \cref{subsubsec: Support Point Sampling Method} (square: integer pixel;
  diamond: sampling centre $\tilde{P}$; dots count toward support, crosses
  do not). (f)~Connect-aware offsetting, realising the offset ratio of
  \cref{eq:p_closed} as moves to lattice neighbours.
  (g)~Topology-aware refinement, insets zooming one guard-kept
  junction source (removal would disconnect) and one removed non-champion
  fragment. (h)~DT-ordered
  thinning (halo: removed width). (i)~The final skeleton; zoom insets show
  raw pixels on the lattice.}
  \label{fig:pipeline_overview}
\end{figure*}

\subsection{Feature-aware Envelope Calculation}\label{subsec: Feature-aware Envelope Calculation}

Separable 1D envelope scans~\cite{meijster2000general} compute the exact EDT
and implicitly map each interior grid index to a minimising boundary; we
retain those minimisers in a deterministic working set. For scan indices
$i,u\in\mathbb{Z}_{\ge 0}$ with $u>i$ the separation coordinate is
\begin{equation}
\xi(i,u) = \frac{u^2-i^2+g(u)^2-g(i)^2}{2(u-i)},
\end{equation}
with $g(\cdot)$ the distance from the previous dimension's pass
and $\operatorname{Sep}(i,u) = \left\lfloor \xi(i,u)\right\rfloor$. The
standard range boundary $w=\operatorname{Sep}(\cdot)+1$ suppresses ties at
integer $\xi(i,u)$, so we replace $\operatorname{Sep}$ by
\begin{equation}
\operatorname{Sep}^*(i,u)=
\begin{cases}
\left\lceil \xi(i,u)\right\rceil-1, & u>i,\\
0, & u=i,
\end{cases}
\end{equation}
and set $w=\operatorname{Sep}^*(\cdot)+1$; the never-queried $u=i$ branch is
fixed for definiteness. Adjacent ranges then overlap at an
integer tie, exposing every tied minimiser at query time while the distance
stays the exact minimum.

Tie discovery is uncapped and the angular span is
computed from the complete tied set in one streaming pass; only sub-pixel
interpolation is bounded, to $N_{\max}$ representative sites per foreground
pixel. That bound is an approximation, not a full-set equivalence, and its
sensitivity is reported with the ablation
(\cref{subsubsec:component_substitution}) and in Supplementary
Appendix~\ref{app:threshold_search}; update rules and complexity are in Supplementary
Appendix~\ref{app:envelope}.

\subsection{Geometry-adaptive Neighbourhood}\label{subsec: Geometry-adaptive Neighbourhood}
Using the digital-extension constructs of \cref{subsec: Discrete Geometry-adaptive Extend Vectors}, we define the geometry-adaptive neighbourhood $\mathcal{P}_Q^P$ for an interior point $P=(p_i)_{i=1}^n \in \mathbb{Z}^n$ and a reference tangent point $Q=(q_i)_{i=1}^n \in \mathbb{Z}^n$. Write $[n]=\{1,\dots,n\}$, let $\mathbb{1}_i$ denote the $i$-th standard basis
vector of $\mathbb{Z}^n$, and, for each $i\in[n]$, set
$\delta_i = p_i - q_i$, $\sigma_i = \operatorname{sgn}(\delta_i)$, and
$\delta_{\max} = \max_{i \in [n]} |\delta_i|$.
Define the main- and secondary-axis index sets $I_m, I_s$ together with their extension vector sets $L_m, L_s$ by
\begin{equation}
\begin{aligned}
I_m = \left\{i \in [n] \mid | \delta_i | = \delta_{\max}\right\},\quad L_m = \left\{\sigma_i \mathbb{1}_i \mid i \in I_m\right\};\\
I_s = \left\{i \in [n] \mid | \delta_i | \in (0,\delta_{\max})\right\},\quad L_s = \left\{\sigma_i \mathbb{1}_i \mid i \in I_s\right\}.
\end{aligned}
\end{equation}

The stepped subset of the geometry-adaptive neighbourhood is
\begin{equation}
\mathcal{P}_s = \bigg\{P + \sum_{l \in L_m} l + \sum_{l \in L_s^{'}} l \,\bigg|\, L_s^{'} \in 2^{L_s}\bigg\},
\end{equation}
stepping along \emph{every} main axis simultaneously (a tied diagonal
displacement yields the single diagonal extension cell).

For axes $i$ with $\delta_i = 0$, collected in $I_o = \{i \in [n] \mid \delta_i = 0\}$, the sign
carries no direction, so the construction probes $Q$'s two axis-neighbours
against the discrete boundary ring, keeping at most one per axis:
$L_o = \{\, s_i\,\mathbb{1}_i \mid i \in I_o,\ s_i \neq 0\}$, where $D$ is the
working index domain, $D^{\circ}=D\setminus\partial\Omega$,
$Q^{\pm}_i=Q\pm\mathbb{1}_i$ and
\begin{equation}
s_i = \begin{cases}
+1, & Q^{+}_i \in D^{\circ},\\
-1, & Q^{+}_i \notin D^{\circ} \text{ and } Q^{-}_i \in D^{\circ},\\
0, & \text{otherwise}
\end{cases}
\end{equation}
so the positive direction wins if both probes qualify. We retain only in-domain foreground
extensions, so
\begin{equation}\label{eq:ga_neighbourhood}
\mathcal{P}^P_Q = \bigg\{\bm{p_s}+\sum_{l \in L_o^{'}} l \;\bigg|\; \bm{p_s} \in \mathcal{P}_s,\, L_o^{'} \in 2^{L_o}\bigg\}\cap D\cap\Omega.
\end{equation}

\subsection{Nearest-shell Multi-tangent Selection}\label{subsec: Geometry-adaptive Neighbourhood Feature Aggregation}

Stage 3 mitigates grid misalignment: each interior grid point gains sub-pixel
candidates along its geometry-adaptive extensions, and at each the span is
read over the nearest sites.

\paragraph{Two-site candidate on an extension}
Let $P\in\Omega\cap\mathbb{Z}^n$ be an interior grid point, $Q$ a reference
tangent point, and $N\in\mathcal{P}^{P}_{Q}$ an extended neighbour
(\cref{subsec: Geometry-adaptive Neighbourhood}). Select tangent sites
$A\in\mathcal{Q}_P$ and $B\in\mathcal{Q}_N$
so that $\|PA\|=d(P)\le \|PB\|$ and $\|NB\|=d(N)\le \|NA\|$, and parameterise
the segment as $\tilde{P}(p)=P+p\,e$ with $e=N-P$ and offset ratio
$p\in[0,1]$.

\begin{lemma}[Equal-distance interpolation]\label{lem:two_site_interp}
Define $\Delta_P = \|PB\|^2-\|PA\|^2$ and
$\Delta_N = \|NA\|^2-\|NB\|^2$.
The selection above gives $\Delta_P\ge 0$ and $\Delta_N\ge 0$; if
$\Delta_P+\Delta_N>0$, the unique $p\in[0,1]$ with
$\|A\tilde{P}(p)\|=\|B\tilde{P}(p)\|$ is
\begin{equation}\label{eq:p_closed}
p \;=\;\frac{\Delta_P}{\Delta_P+\Delta_N}.
\end{equation}
\end{lemma}

Supplementary Appendix~\ref{app:incircle_full} proves it; the pair is equidistant from
$\tilde{P}=P+p\,e$ by construction. The method exports that centre alone,
the integer mark plus the offset $p\,e$; no radius is exported per pixel.

\paragraph{The nearest shell, at $p=0$ and at $p>0$}
Every pixel first receives the grid-point reference
$S_{\mathrm{ref}} = S(P;\mathcal{Q}_P)$ with its pair and support evidence;
that span lies on one distance shell, globally nearest by
\cref{eq:tangent_set}. Stage 3 applies the same rule at $p>0$. Only reference sites whose own stencil reaches $N$ pair, so
with $\mathcal{Q}^{N}_{P}=\{A\in\mathcal{Q}_P \mid N\in\mathcal{P}^{P}_{A}\}$ the
admissible bucket at ratio $p$ is the subset
$G_p\subseteq\mathcal{Q}^{N}_{P}\times\mathcal{Q}_N$ of pairs $(A,B)$ with
$B\neq A$, $\|PB\|\neq d(P)$ and offset ratio $p$. They meet at one
$\tilde{P}=P+p\,e$, and the sites in play are
\begin{equation}
\mathcal{Q}^{(p)} = \pi_1(G_p)\cup\pi_2(G_p),
\end{equation}
the two projections of $G_p$. $\mathcal{Q}_P$ and $\mathcal{Q}_N$ being
shells about the \emph{different} centres $P$ and $N$, they need not stay
co-radial from $\tilde{P}$. Of $\mathcal{Q}^{(p)}$ we keep only the subset
$\mathcal{Q}^{\min}_{\tilde{P}}$ at minimum distance $r_{\ast}$ from it, read
the span $S(\tilde{P};\mathcal{Q}^{\min}_{\tilde{P}})$ on that shell, and
reject candidates with fewer than two. The minimum runs
over $\mathcal{Q}^{(p)}$, not $\partial\Omega$: $r_{\ast}$ is the nearest
\emph{participating} radius, and membership and the
minimum are exact integer comparisons (Supplementary
Appendix~\ref{app:incircle_full}).

The best survivor replaces the reference, giving the $\widehat S_P$
that \cref{subsec: Skeleton Filtering} thresholds, only when its support
evidence is at least as strong and its span strictly larger; pixels with
$S_{\mathrm{ref}}\ge\pi$, or reference sites on different contour components,
skip refinement, and grouping and tie-breaks are in Supplementary
Appendix~\ref{app:incircle_full}. The two sites
realising the span need not be an enumerated pair.
Groups are keyed by the exact rational, never a floating-point value: with
$p_n=\Delta_P$, $p_d=\Delta_P+\Delta_N=2e^{\top}(B-A)$ integers in the grid data
and $g=\gcd(p_n,p_d)$, the key is the reduced pair $(p_n/g,\,p_d/g)$, so two pairs
group together exactly when they meet at one $\tilde{P}$. That fixes \emph{where}
a group sits; it does not make the group co-radial, and the nearest-shell
retention above, not the key, drops the cross-shell sites.

\subsection{Skeleton Filtering}\label{subsec: Skeleton Filtering}

Stage 4 admits or prunes each pixel by $\widehat S_P$.

\subsubsection{Tangent Angle Span Filtering}
The selected span $\widehat S_P$ of Stage 3 is a significance score, not a
verdict on mediality: it is large where the retained tangents surround the
pixel and small where they cluster on a single local arc. Two angular cutoffs
$\phi_l < \phi_h$ partition each pixel $P$ into three regimes:
\begin{equation}\label{eq:phi_thresholds}
\begin{cases}
\widehat S_P \le \phi_l, & P \text{ pruned (low span)};\\
\phi_l < \widehat S_P < \phi_h, & P \text{ retained iff the support test}\\
                       & \text{below passes};\\
\phi_h \le \widehat S_P, & P \text{ retained (high span)}.
\end{cases}
\end{equation}
The criterion inherits the classical significance tradition that prunes medial
points by the angle their nearest boundary contacts
subtend~\cite{DBLP:journals/ivc/MalandainF98, DBLP:conf/sma/FoskeyLM03,
DBLP:journals/ivc/CouprieCZ07}, to which the span reduces exactly in the
generic two-site case. It differs in how that angle is read: over the whole
retained shell rather than one contact pair, at the upgraded sub-pixel
candidate wherever Stage 3 admits one, and through two cutoffs whose middle
band each pixel settles locally by support, not propagation;
$(\phi_l,\phi_h)$ are calibrated once on a discrete grid and reported in
Section~\ref{sec:Experiment}.

\subsubsection{Support Point Sampling Method}\label{subsubsec: Support Point Sampling Method}
In 2D the support test is deterministic. Given $\tilde P$ and the retained
pair $(A^{\ast},B^{\ast})$ realising its span, it rotates $A^{\ast}-\tilde P$
toward $B^{\ast}-\tilde P$ and takes the support radius $r$ from
$r^2=\max(\|A^{\ast}-\tilde P\|^2,\|B^{\ast}-\tilde P\|^2)$; the retained
sites are co-radial, so $r=r_{\ast}$ exactly. We sample
$N_{\mathrm{s}}=15$
equally spaced points
$A_t$ on the shorter arc between $A^{\ast}$ and $B^{\ast}$, let
$\vec{\bm e}_t$ be the direction of $A_t$ from $\tilde P$ scaled to
unit Euclidean norm, and probe
$Z_t = \operatorname{round}\!\left(A_t+k(\varphi,r)\vec{\bm e}_t\right)$
for $t=1,\dots,N_{\mathrm{s}}$, where $\varphi$ is the observed tangent span,
$r$ is the support radius above, $\phi_d=\pi/2$, and
\begin{equation}
k(\varphi,r)=
\begin{cases}
1, & \varphi<\phi_d,\\
(\pi-\varphi)/(\pi-\phi_d), & \varphi\ge\phi_d,\ r^2<4,\\
2(\pi-\varphi)/(\pi-\phi_d)-1, & \varphi\ge\phi_d,\ r^2\ge4.
\end{cases}
\end{equation}
For $r^2\ge4$, $k$ crosses zero at $\varphi=3\pi/4$ and is negative up to
$\phi_h$: high spans probe inward toward $\tilde P$, and at the crossing
probe and arc point coincide, reading one sample.
Coordinates are clipped to the image. A sample counts only if the rounded arc
point and probe are in the same foreground component as the current pixel,
neither coincides with either tangent, and the probe does not coincide with
the current pixel. The support flag is
true iff at least one sample counts. Tangents on different contour components
pass immediately, as do spans above $\pi$; the cutoffs of
\cref{eq:phi_thresholds} act only in Stage 4.

\subsection{Algorithm Pipeline}\label{subsec: Algorithm Pipeline}

Let $B$ denote the input binary mask, $\mathcal A_{\phi_l,\phi_h}(B)$ the six-stage 2D extractor, and
$\mathcal R_{\tau_{\max}}(S;B)$ the optional closing reconstruction-aware
reduction. The three SpiderSkel variants are the extractor itself,
\begin{equation}
S=\mathcal A_{\phi_l,\phi_h}(B),
\qquad
S^{+R_2}=\mathcal R_{2}\!\left(\mathcal A_{\phi_l,\phi_h}(B);B\right),
\label{eq:complete_method}
\end{equation}
instantiated in three variants. All reported results use
$(\phi_l,\phi_h)=(0,13\pi/16)$, selected on the evaluation corpus
(Section~\ref{sec:Experiment}) and fixed for every downstream evaluation with no
per-dataset retuning, except where a sensitivity or ablation study explicitly
varies it; the optional reduction is applied in no reported result, and its own
measurements accompany the released package.

The extractor applies its six stages in this order, as SpiderSkel-N8 realises
them; SpiderSkel-DualC substitutes the Stage-3/4 support read and the three
tails differ, both specified below.
(1)~\emph{Envelope EDT} (\cref{subsec: Feature-aware Envelope Calculation}):
the exact $d(P)$, complete tie discovery, $\le N_{\max}$ representatives.
(2)~\emph{Geometry-adaptive neighbourhood}
(\cref{subsec: Geometry-adaptive Neighbourhood}): the extension stencil.
(3)~\emph{Nearest-shell multi-tangent selection}
(\cref{subsec: Geometry-adaptive Neighbourhood Feature Aggregation}): the
sub-grid centre $\tilde P$ and the span on its nearest participating shell.
(4)~\emph{Span admission}: the cutoffs of \cref{eq:phi_thresholds} on
$\widehat S_P$, the borderline band decided by the support test
(\cref{subsubsec: Support Point Sampling Method}).
(5)~\emph{Connect-aware offsetting}, candidates in row-major order on one
evolving set: the ratio $p$ of \cref{eq:p_closed}
proposes the lattice endpoint $P+e$ by default when $p>\tfrac12$; a
\emph{mutually pointing} candidate, whose target proposes the reverse offset
with both ratios at least $\tfrac12$, instead moves iff $A^{\ast}$ follows
$B^{\ast}$ in row-major order in its selected pair. The add guard writes the
target only when, on the current set, exactly one $4$-connected background
component of its $3\times3$ ring is axially adjacent to it (the hole-count
half of $(8,4)$-simplicity); otherwise neither pixel changes. The source is
then vacated only where it has taken no earlier mover's target and is
$(8,4)$-simple in the current set.
(6)~\emph{Topology-aware refinement and thinning}, specified below.
The output mask stays on the integer lattice; the sub-pixel centres
$\tilde P$ feed only the sub-pixel decodes of Section~\ref{sec:Experiment},
whose radii, like the integer decode's, are recomputed from the object's own
distance transform, never from $r_{\ast}$.
Fig.~\ref{fig:pipeline_overview} traces all six on one exemplar.

SpiderSkel-N8 and SpiderSkel-Scaffold share one evidence front: stages 1--4
are identical. SpiderSkel-DualC changes that front at the
support read of stages 3 and 4, reading the foreground's $8$-connected object
labels and $4$-connected part structure jointly, so a pinch certificate adds
same-object support mass across a pinch that the ordinary read rejects, never
the reverse (within a radius-$4$ window, the read cell is $8$-reachable but
not $4$-reachable from the candidate over same-object pixels); that bit leads
the Stage-3 selection key and decides the Stage-4 band. The tails then
differ.
SpiderSkel-N8 partitions each object into $4$-connected parts, applies
the connect-aware move, keeps one champion fragment per part, adds
the constructive certificate fills, and closes with DT-ordered thinning.
SpiderSkel-DualC keeps that move and that thinning but cleans per shape and
substitutes a single constructive cycle fill. SpiderSkel-Scaffold instead
anchors every surviving candidate (movers at their offset targets, the rest in
place), seeding any anchor-free $4$-connected part at its deepest-EDT pixel
so it survives, and erodes the object mask down onto that scaffold by $(8,4)$ simple-point
deletion alone, in ascending EDT order, in place of the move and the refinement, and closes with
coverage-guarded thinning, whose extra test only withholds deletions: none may
strand object mass outside every surviving reconstruction disc. Every edit is
such a deletion and this variant writes no pixel, so each object's component and
cycle counts are preserved by construction, a claim about those counts, not a
homotopy equivalence. For SpiderSkel-N8 and SpiderSkel-DualC the
certificates are the evidence.

Stage 6, as SpiderSkel-N8 runs it, includes interior filling, cleaning to one
champion fragment per 4-connected input component, and DT-ordered,
coverage-guarded thinning, with every pixel sweep in row-major order;
Supplementary Appendix~\ref{app:closing_stages} states each rule, the two
other tails, and the closing $\mathcal R_2$ below.

The optional closing $\mathcal R_2$ deletes only skeleton pixels whose cyclic
8-neighbourhood has one $0\!\to\!1$ transition and whose reconstruction disc
uniquely covers at most $\tau$ pixels, sweeping $\tau$ upward while the
clipped reconstruction keeps the input's 8-connected Euler number. It is the
optional $+R_2$ enhancement, measured separately in the released package. All of these are local
properties; whether the six-stage
construction is homotopy-equivalent to the input remains open.
 
\section{Experiment}\label{sec:Experiment}
\subsection{Experimental Setup}

\subsubsection{Datasets}

Four real-world domains form the main benchmark of nineteen datasets.
\textbf{Shape matching} (12):
Animal2000~\cite{DBLP:conf/iccvw/BaiLT09},
Articulated Shapes~\cite{DBLP:journals/pami/LingJ07},
EM200~\cite{DBLP:conf/icpr/YangLTSG14},
Kimia216~\cite{DBLP:journals/pami/SebastianKK04},
Kimia99~\cite{DBLP:journals/pami/SebastianKK04},
MPEG400~\cite{DBLP:conf/icip/YangTPFKG14},
MPEG7~\cite{DBLP:conf/cvpr/LateckiLE00},
Smithsonian Leaves~\cite{DBLP:journals/pami/LingJ07},
Swedish Leaves~\cite{soderkvist2001computer},
Tari56~\cite{DBLP:conf/iccv/AslanT05},
Tetrapod120~\cite{DBLP:journals/pr/YangTSG16},
and WH-SYMMAX~\cite{DBLP:journals/pr/ShenBHZ16};
\textbf{retinal vessel} (3):
CHASE\_DB1~\cite{DBLP:journals/tbe/FrazRHUROB12},
DRIVE~\cite{DBLP:journals/tmi/StaalANVG04},
and STARE~\cite{hoover2000locating,DBLP:journals/tmi/HooverG03};
\textbf{crack and road} (3):
CrackForest~\cite{DBLP:journals/tits/ShiCQMC16,DBLP:conf/icdsci/CuiQCMS15},
DeepGlobe~\cite{DBLP:conf/cvpr/DemirKLPHBHTR18},
and Massachusetts Roads~\cite{DBLP:phd/ca/Mnih13};
and \textbf{document binarisation} (1): the pooled DIBCO/H-DIBCO 2009--2018
competition sets~\cite{DBLP:conf/icdar/GatosNP09,
  DBLP:conf/icfhr/PratikakisGN10,
  DBLP:conf/icdar/PratikakisGN11,
  DBLP:conf/icfhr/PratikakisGN12,
  DBLP:conf/icdar/PratikakisGN13,
  DBLP:conf/icfhr/NtirogiannisGP14,
  DBLP:conf/icfhr/PratikakisZBG16,
  DBLP:conf/icdar/PratikakisZBG17,
  DBLP:conf/icfhr/PratikakisZKG18}.
Shape benchmarks take binary object masks; the other three domains, ground-truth foreground maps.

\subsubsection{Compared Methods}
We compare the three variants against twenty-two baselines spanning
every method family, cited in
Table~\ref{tab:category_avg_skelqs_em}. SpiderSkel runs without the optional
reduction throughout the headline benchmark, and the learned models are
evaluated as released (configurations: Supplementary
Table~\ref{tab:baseline_configuration}); Supplementary
Appendix~\ref{app:learned_comparison} retrains both
on reference annotations for architecture-fair comparison.

\begin{table*}[!htb]
\centering
\footnotesize
\setlength{\tabcolsep}{3pt}
\renewcommand{\arraystretch}{1.00}
\caption{Category-wise comparison, scores in \%, higher is better. SRQS is
  SkelReconQS, the composite skeleton-reconstruction quality score
  (\cref{eq:skelqs}); Reduct.\ is the Reduction Rate, the fraction of
  foreground the skeleton discards; Dice$_{\mathrm{sp}}$ is strict sub-pixel
  reconstruction Dice (Section~\ref{subsubsec:srqs_metric} defines all three).
  SRQS cells additionally stack $T_{\mathrm{skel}}$\,/\,$T_{\mathrm{recon}}$,
  the skeleton-topology and reconstruction-topology certificates, on a $[0,1]$
  scale; a bold green
  \textcolor[HTML]{1B7837}{\textbf{1.0}} denotes exactly one. The top three
  SRQS entries per column carry a green gradient (\textbf{best} bold,
  \underline{second} underlined); Reduct.\ and Dice$_{\mathrm{sp}}$ are
  unranked. Overall is the equal-weight macro-mean over the nineteen
  datasets, $(\pm h)$ its 95\% confidence half-width; stars mark
  Holm-corrected significance against the highest-scoring baseline in the
  Overall column ($^{*}p{<}0.05$, $^{**}p{<}0.01$). Ties break at full precision (SpiderSkel-N8 and SpiderSkel-DualC, $0.004$~pp apart in Overall, are effectively tied); implementations: Supplementary Table~\ref{tab:baseline_configuration}}
\label{tab:category_avg_skelqs_em}
\setlength{\tabcolsep}{2.3pt}%
\begin{tabular}{@{}c l *{4}{c c c} !{\vrule width 0.4pt}c@{}}
\toprule
  &        & \multicolumn{12}{c}{\textit{Per category}}                                  & \textit{Aggregate} \\
\cmidrule(lr){3-14} \cmidrule(lr){15-15}
  &        & \multicolumn{3}{c}{Shape matching} & \multicolumn{3}{c}{Retinal vessel} & \multicolumn{3}{c}{Crack \& road} & \multicolumn{3}{c}{Document bin.} &  \\
\cmidrule(lr){3-5} \cmidrule(lr){6-8} \cmidrule(lr){9-11} \cmidrule(lr){12-14}
  & Method & {\scriptsize SRQS$\uparrow$} & {\scriptsize Reduct.$\uparrow$} & {\scriptsize Dice$_{\mathrm{sp}}\uparrow$} & {\scriptsize SRQS$\uparrow$} & {\scriptsize Reduct.$\uparrow$} & {\scriptsize Dice$_{\mathrm{sp}}\uparrow$} & {\scriptsize SRQS$\uparrow$} & {\scriptsize Reduct.$\uparrow$} & {\scriptsize Dice$_{\mathrm{sp}}\uparrow$} & {\scriptsize SRQS$\uparrow$} & {\scriptsize Reduct.$\uparrow$} & {\scriptsize Dice$_{\mathrm{sp}}\uparrow$} & Overall$\uparrow$ \\
\midrule
\multirow{3}{*}[-0.6em]{\raisebox{0pt}[0pt][0pt]{\rotatebox[origin=c]{90}{\scriptsize\itshape SpiderSkel}}} & \cellcolor{black!8}\textbf{N8} & \cellcolor{black!8}\stkbest{96.47}{\textcolor[HTML]{1B7837}{\textbf{1.0}}\,/\,\textcolor[HTML]{1B7837}{\textbf{1.0}}} & \cellcolor{black!8}\stkcell{95.71}{\vphantom{1.0}} & \cellcolor{black!8}\stkcell{99.82}{\vphantom{1.0}} & \cellcolor{black!8}\stksec{79.71}{\textcolor[HTML]{1B7837}{\textbf{1.0}}\,/\,\textcolor[HTML]{1B7837}{\textbf{1.0}}} & \cellcolor{black!8}\stkcell{76.15}{\vphantom{1.0}} & \cellcolor{black!8}\stkcell{99.51}{\vphantom{1.0}} & \cellcolor{black!8}\stksec{83.68}{\textcolor[HTML]{1B7837}{\textbf{1.0}}\,/\,\textcolor[HTML]{1B7837}{\textbf{1.0}}} & \cellcolor{black!8}\stkcell{82.42}{\vphantom{1.0}} & \cellcolor{black!8}\stkcell{99.91}{\vphantom{1.0}} & \cellcolor{black!8}\stkbest{82.89}{\textcolor[HTML]{1B7837}{\textbf{1.0}}\,/\,\textcolor[HTML]{1B7837}{\textbf{1.0}}} & \cellcolor{black!8}\stkcell{82.13}{\vphantom{1.0}} & \cellcolor{black!8}\stkcell{99.83}{\vphantom{1.0}} & \cellcolor{black!8}\stkbest{91.09$^{**}$\,{\scriptsize($\pm$0.11)}}{\textcolor[HTML]{1B7837}{\textbf{1.0}}\,/\,\textcolor[HTML]{1B7837}{\textbf{1.0}}} \\
 & \cellcolor{black!8}\textbf{DualC} & \cellcolor{black!8}\stksec{96.47}{\textcolor[HTML]{1B7837}{\textbf{1.0}}\,/\,\textcolor[HTML]{1B7837}{\textbf{1.0}}} & \cellcolor{black!8}\stkcell{95.70}{\vphantom{1.0}} & \cellcolor{black!8}\stkcell{99.82}{\vphantom{1.0}} & \cellcolor{black!8}\stkthird{79.70}{\textcolor[HTML]{1B7837}{\textbf{1.0}}\,/\,\textcolor[HTML]{1B7837}{\textbf{1.0}}} & \cellcolor{black!8}\stkcell{76.12}{\vphantom{1.0}} & \cellcolor{black!8}\stkcell{99.51}{\vphantom{1.0}} & \cellcolor{black!8}\stkthird{83.68}{\textcolor[HTML]{1B7837}{\textbf{1.0}}\,/\,\textcolor[HTML]{1B7837}{\textbf{1.0}}} & \cellcolor{black!8}\stkcell{82.41}{\vphantom{1.0}} & \cellcolor{black!8}\stkcell{99.91}{\vphantom{1.0}} & \cellcolor{black!8}\stksec{82.88}{\textcolor[HTML]{1B7837}{\textbf{1.0}}\,/\,\textcolor[HTML]{1B7837}{\textbf{1.0}}} & \cellcolor{black!8}\stkcell{82.11}{\vphantom{1.0}} & \cellcolor{black!8}\stkcell{99.83}{\vphantom{1.0}} & \cellcolor{black!8}\stksec{91.09$^{**}$\,{\scriptsize($\pm$0.11)}}{\textcolor[HTML]{1B7837}{\textbf{1.0}}\,/\,\textcolor[HTML]{1B7837}{\textbf{1.0}}} \\
 & \cellcolor{black!8}\textbf{Scaffold} & \cellcolor{black!8}\stkthird{96.22}{\textcolor[HTML]{1B7837}{\textbf{1.0}}\,/\,\textcolor[HTML]{1B7837}{\textbf{1.0}}} & \cellcolor{black!8}\stkcell{94.91}{\vphantom{1.0}} & \cellcolor{black!8}\stkcell{99.87}{\vphantom{1.0}} & \cellcolor{black!8}\stkbest{79.73}{\textcolor[HTML]{1B7837}{\textbf{1.0}}\,/\,\textcolor[HTML]{1B7837}{\textbf{1.0}}} & \cellcolor{black!8}\stkcell{75.68}{\vphantom{1.0}} & \cellcolor{black!8}\stkcell{99.51}{\vphantom{1.0}} & \cellcolor{black!8}\stkbest{83.79}{\textcolor[HTML]{1B7837}{\textbf{1.0}}\,/\,\textcolor[HTML]{1B7837}{\textbf{1.0}}} & \cellcolor{black!8}\stkcell{82.12}{\vphantom{1.0}} & \cellcolor{black!8}\stkcell{99.92}{\vphantom{1.0}} & \cellcolor{black!8}\stkthird{82.64}{\textcolor[HTML]{1B7837}{\textbf{1.0}}\,/\,\textcolor[HTML]{1B7837}{\textbf{1.0}}} & \cellcolor{black!8}\stkcell{81.31}{\vphantom{1.0}} & \cellcolor{black!8}\stkcell{99.91}{\vphantom{1.0}} & \cellcolor{black!8}\stkthird{90.94$^{**}$\,{\scriptsize($\pm$0.11)}}{\textcolor[HTML]{1B7837}{\textbf{1.0}}\,/\,\textcolor[HTML]{1B7837}{\textbf{1.0}}} \\
\midrule
\multirow{9}{*}{\raisebox{0pt}[0pt][0pt]{\rotatebox[origin=c]{90}{\scriptsize\itshape Distance transform}}} & medial\_axis~\cite{blum1967MAT} & \stkcell{92.18}{0.65\,/\,\textcolor[HTML]{1B7837}{\textbf{1.0}}} & \stkcell{94.90}{\vphantom{1.0}} & \stkcell{99.23}{\vphantom{1.0}} & \stkcell{76.98}{0.26\,/\,\textcolor[HTML]{1B7837}{\textbf{1.0}}} & \stkcell{75.74}{\vphantom{1.0}} & \stkcell{97.29}{\vphantom{1.0}} & \stkcell{82.46}{0.62\,/\,\textcolor[HTML]{1B7837}{\textbf{1.0}}} & \stkcell{82.09}{\vphantom{1.0}} & \stkcell{98.43}{\vphantom{1.0}} & \stkcell{61.38}{0.21\,/\,0.99} & \stkcell{82.33}{\vphantom{1.0}} & \stkcell{97.56}{\vphantom{1.0}} & \stkcell{86.62\,{\scriptsize($\pm$0.30)}}{0.56\,/\,0.99} \\
 & hjs~\cite{DBLP:journals/ijcv/SiddiqiBTZ02} & \stkcell{95.60}{\textcolor[HTML]{1B7837}{\textbf{1.0}}\,/\,\textcolor[HTML]{1B7837}{\textbf{1.0}}} & \stkcell{96.06}{\vphantom{1.0}} & \stkcell{99.49}{\vphantom{1.0}} & \stkcell{32.33}{\textcolor[HTML]{1B7837}{\textbf{1.0}}\,/\,\textcolor[HTML]{1B7837}{\textbf{1.0}}} & \stkcell{76.76}{\vphantom{1.0}} & \stkcell{96.91}{\vphantom{1.0}} & \stkcell{79.31}{\textcolor[HTML]{1B7837}{\textbf{1.0}}\,/\,\textcolor[HTML]{1B7837}{\textbf{1.0}}} & \stkcell{82.62}{\vphantom{1.0}} & \stkcell{98.06}{\vphantom{1.0}} & \stkcell{74.58}{\textcolor[HTML]{1B7837}{\textbf{1.0}}\,/\,0.99} & \stkcell{82.62}{\vphantom{1.0}} & \stkcell{98.03}{\vphantom{1.0}} & \stkcell{81.93\,{\scriptsize($\pm$0.71)}}{\textcolor[HTML]{1B7837}{\textbf{1.0}}\,/\,0.99} \\
 & dsepruning~\cite{DBLP:journals/pami/BaiLL07} & \stkcell{81.69}{0.95\,/\,0.99} & \stkcell{96.94}{\vphantom{1.0}} & \stkcell{97.47}{\vphantom{1.0}} & \stkcell{5.43}{0.22\,/\,0.40} & \stkcell{78.47}{\vphantom{1.0}} & \stkcell{95.35}{\vphantom{1.0}} & \stkcell{38.74}{0.59\,/\,0.70} & \stkcell{82.87}{\vphantom{1.0}} & \stkcell{97.67}{\vphantom{1.0}} & \stkcell{0.11}{0.00\,/\,0.03} & \stkcell{86.71}{\vphantom{1.0}} & \stkcell{86.86}{\vphantom{1.0}} & \stkcell{58.57\,{\scriptsize($\pm$0.73)}}{0.73\,/\,0.80} \\
 & voronoi~\cite{voronoi1908nouvelles} & \stkcell{92.91}{0.88\,/\,0.99} & \stkcell{96.36}{\vphantom{1.0}} & \stkcell{99.35}{\vphantom{1.0}} & \stkcell{6.60}{0.02\,/\,0.23} & \stkcell{78.23}{\vphantom{1.0}} & \stkcell{95.78}{\vphantom{1.0}} & \stkcell{55.61}{0.42\,/\,0.73} & \stkcell{82.50}{\vphantom{1.0}} & \stkcell{97.89}{\vphantom{1.0}} & \stkcell{35.37}{0.06\,/\,0.46} & \stkcell{82.32}{\vphantom{1.0}} & \stkcell{97.65}{\vphantom{1.0}} & \stkcell{70.37\,{\scriptsize($\pm$0.69)}}{0.63\,/\,0.80} \\
 & matlab\_voronoi~\cite{yohai_voronoi_fileexchange_citeas} & \stkcell{50.32}{0.33\,/\,0.99} & \stkcell{95.17}{\vphantom{1.0}} & \stkcell{98.89}{\vphantom{1.0}} & \stkcell{6.12}{0.00\,/\,0.20} & \stkcell{75.95}{\vphantom{1.0}} & \stkcell{92.22}{\vphantom{1.0}} & \stkcell{46.13}{0.22\,/\,0.73} & \stkcell{80.20}{\vphantom{1.0}} & \stkcell{94.26}{\vphantom{1.0}} & \stkcell{44.00}{0.00\,/\,0.66} & \stkcell{76.85}{\vphantom{1.0}} & \stkcell{94.81}{\vphantom{1.0}} & \stkcell{42.35\,{\scriptsize($\pm$1.09)}}{0.24\,/\,0.80} \\
 & aof~\cite{rezanejad2013flux,DBLP:conf/cvpr/DimitrovDS03} & \stkcell{12.79}{0.02\,/\,0.97} & \stkcell{96.32}{\vphantom{1.0}} & \stkcell{99.57}{\vphantom{1.0}} & \stkcell{1.53}{0.00\,/\,0.11} & \stkcell{83.24}{\vphantom{1.0}} & \stkcell{92.11}{\vphantom{1.0}} & \stkcell{44.98}{0.17\,/\,0.67} & \stkcell{84.35}{\vphantom{1.0}} & \stkcell{96.11}{\vphantom{1.0}} & \stkcell{32.03}{0.00\,/\,0.69} & \stkcell{82.86}{\vphantom{1.0}} & \stkcell{96.94}{\vphantom{1.0}} & \stkcell{17.11\,{\scriptsize($\pm$0.57)}}{0.04\,/\,0.77} \\
 & afmm+tft~\cite{DBLP:conf/vissym/TeleaW02,DBLP:conf/visapp/ReniersT06a} & \stkcell{60.99}{0.58\,/\,0.86} & \stkcell{97.32}{\vphantom{1.0}} & \stkcell{96.13}{\vphantom{1.0}} & \stkcell{0.64}{0.01\,/\,0.12} & \stkcell{83.59}{\vphantom{1.0}} & \stkcell{88.24}{\vphantom{1.0}} & \stkcell{10.68}{0.22\,/\,0.38} & \stkcell{86.19}{\vphantom{1.0}} & \stkcell{90.53}{\vphantom{1.0}} & \stkcell{0.00}{0.00\,/\,0.00} & \stkcell{88.15}{\vphantom{1.0}} & \stkcell{79.28}{\vphantom{1.0}} & \stkcell{40.31\,{\scriptsize($\pm$0.85)}}{0.40\,/\,0.62} \\
 & afmm~\cite{DBLP:conf/vissym/TeleaW02} & \stkcell{31.09}{0.07\,/\,\textcolor[HTML]{1B7837}{\textbf{1.0}}} & \stkcell{91.76}{\vphantom{1.0}} & \stkcell{99.78}{\vphantom{1.0}} & \stkcell{48.07}{0.30\,/\,0.72} & \stkcell{76.40}{\vphantom{1.0}} & \stkcell{96.81}{\vphantom{1.0}} & \stkcell{73.89}{0.70\,/\,0.91} & \stkcell{82.55}{\vphantom{1.0}} & \stkcell{97.98}{\vphantom{1.0}} & \stkcell{61.90}{0.36\,/\,0.81} & \stkcell{82.03}{\vphantom{1.0}} & \stkcell{97.67}{\vphantom{1.0}} & \stkcell{42.15\,{\scriptsize($\pm$1.19)}}{0.22\,/\,0.93} \\
 & kimimaro~\cite{DBLP:conf/pg/SatoBBKN00} & \stkcell{48.18}{0.99\,/\,0.99} & \stkcell{97.91}{\vphantom{1.0}} & \stkcell{88.97}{\vphantom{1.0}} & \stkcell{0.00}{0.00\,/\,0.62} & \stkcell{95.35}{\vphantom{1.0}} & \stkcell{43.62}{\vphantom{1.0}} & \stkcell{17.80}{0.47\,/\,0.79} & \stkcell{86.78}{\vphantom{1.0}} & \stkcell{79.29}{\vphantom{1.0}} & \stkcell{1.87}{0.00\,/\,0.56} & \stkcell{88.36}{\vphantom{1.0}} & \stkcell{80.45}{\vphantom{1.0}} & \stkcell{33.34\,{\scriptsize($\pm$0.77)}}{0.70\,/\,0.88} \\
\midrule
\multirow{11}{*}{\raisebox{0pt}[0pt][0pt]{\rotatebox[origin=c]{90}{\scriptsize\itshape Iterative thinning}}} & zhang~\cite{DBLP:journals/cacm/ZhangS84} & \stkcell{89.46}{\textcolor[HTML]{1B7837}{\textbf{1.0}}\,/\,0.99} & \stkcell{96.29}{\vphantom{1.0}} & \stkcell{97.90}{\vphantom{1.0}} & \stkcell{70.26}{\textcolor[HTML]{1B7837}{\textbf{1.0}}\,/\,0.98} & \stkcell{76.22}{\vphantom{1.0}} & \stkcell{96.62}{\vphantom{1.0}} & \stkcell{79.57}{\textcolor[HTML]{1B7837}{\textbf{1.0}}\,/\,0.99} & \stkcell{82.59}{\vphantom{1.0}} & \stkcell{97.64}{\vphantom{1.0}} & \stkcell{68.40}{\textcolor[HTML]{1B7837}{\textbf{1.0}}\,/\,0.99} & \stkcell{82.97}{\vphantom{1.0}} & \stkcell{96.56}{\vphantom{1.0}} & \stkcell{83.76\,{\scriptsize($\pm$0.40)}}{\textcolor[HTML]{1B7837}{\textbf{1.0}}\,/\,0.99} \\
 & guo-hall~\cite{DBLP:journals/cacm/GuoH89} & \stkcell{89.04}{\textcolor[HTML]{1B7837}{\textbf{1.0}}\,/\,\textcolor[HTML]{1B7837}{\textbf{1.0}}} & \stkcell{96.33}{\vphantom{1.0}} & \stkcell{97.69}{\vphantom{1.0}} & \stkcell{73.11}{\textcolor[HTML]{1B7837}{\textbf{1.0}}\,/\,0.98} & \stkcell{76.38}{\vphantom{1.0}} & \stkcell{96.37}{\vphantom{1.0}} & \stkcell{79.72}{\textcolor[HTML]{1B7837}{\textbf{1.0}}\,/\,0.99} & \stkcell{82.65}{\vphantom{1.0}} & \stkcell{97.62}{\vphantom{1.0}} & \stkcell{71.10}{\textcolor[HTML]{1B7837}{\textbf{1.0}}\,/\,0.99} & \stkcell{83.11}{\vphantom{1.0}} & \stkcell{96.55}{\vphantom{1.0}} & \stkcell{84.11\,{\scriptsize($\pm$0.41)}}{\textcolor[HTML]{1B7837}{\textbf{1.0}}\,/\,0.99} \\
 & thinning~\cite{DBLP:journals/cacm/GuoH89} & \stkcell{88.91}{\textcolor[HTML]{1B7837}{\textbf{1.0}}\,/\,\textcolor[HTML]{1B7837}{\textbf{1.0}}} & \stkcell{96.33}{\vphantom{1.0}} & \stkcell{97.67}{\vphantom{1.0}} & \stkcell{73.41}{\textcolor[HTML]{1B7837}{\textbf{1.0}}\,/\,0.98} & \stkcell{76.38}{\vphantom{1.0}} & \stkcell{96.36}{\vphantom{1.0}} & \stkcell{79.36}{\textcolor[HTML]{1B7837}{\textbf{1.0}}\,/\,0.99} & \stkcell{82.66}{\vphantom{1.0}} & \stkcell{97.62}{\vphantom{1.0}} & \stkcell{69.94}{\textcolor[HTML]{1B7837}{\textbf{1.0}}\,/\,0.99} & \stkcell{83.11}{\vphantom{1.0}} & \stkcell{96.65}{\vphantom{1.0}} & \stkcell{83.96\,{\scriptsize($\pm$0.39)}}{\textcolor[HTML]{1B7837}{\textbf{1.0}}\,/\,0.99} \\
 & luwang~\cite{DBLP:journals/cacm/LuW86} & \stkcell{88.34}{\textcolor[HTML]{1B7837}{\textbf{1.0}}\,/\,0.99} & \stkcell{95.89}{\vphantom{1.0}} & \stkcell{97.74}{\vphantom{1.0}} & \stkcell{71.31}{\textcolor[HTML]{1B7837}{\textbf{1.0}}\,/\,0.98} & \stkcell{72.74}{\vphantom{1.0}} & \stkcell{97.95}{\vphantom{1.0}} & \stkcell{77.79}{0.99\,/\,0.99} & \stkcell{81.08}{\vphantom{1.0}} & \stkcell{98.05}{\vphantom{1.0}} & \stkcell{46.58}{0.67\,/\,0.66} & \stkcell{80.65}{\vphantom{1.0}} & \stkcell{97.52}{\vphantom{1.0}} & \stkcell{81.79\,{\scriptsize($\pm$0.47)}}{0.98\,/\,0.97} \\
 & lee~\cite{DBLP:journals/cvgip/LeeKC94} & \stkcell{83.16}{\textcolor[HTML]{1B7837}{\textbf{1.0}}\,/\,0.99} & \stkcell{96.81}{\vphantom{1.0}} & \stkcell{96.37}{\vphantom{1.0}} & \stkcell{64.90}{\textcolor[HTML]{1B7837}{\textbf{1.0}}\,/\,\textcolor[HTML]{1B7837}{\textbf{1.0}}} & \stkcell{76.42}{\vphantom{1.0}} & \stkcell{96.54}{\vphantom{1.0}} & \stkcell{72.34}{\textcolor[HTML]{1B7837}{\textbf{1.0}}\,/\,0.99} & \stkcell{82.76}{\vphantom{1.0}} & \stkcell{97.66}{\vphantom{1.0}} & \stkcell{56.71}{\textcolor[HTML]{1B7837}{\textbf{1.0}}\,/\,0.99} & \stkcell{83.67}{\vphantom{1.0}} & \stkcell{95.88}{\vphantom{1.0}} & \stkcell{77.18\,{\scriptsize($\pm$0.37)}}{\textcolor[HTML]{1B7837}{\textbf{1.0}}\,/\,0.99} \\
 & zhangsuen~\cite{DBLP:journals/cacm/ZhangS84} & \stkcell{85.50}{\textcolor[HTML]{1B7837}{\textbf{1.0}}\,/\,0.99} & \stkcell{96.31}{\vphantom{1.0}} & \stkcell{96.89}{\vphantom{1.0}} & \stkcell{43.85}{\textcolor[HTML]{1B7837}{\textbf{1.0}}\,/\,0.98} & \stkcell{73.00}{\vphantom{1.0}} & \stkcell{97.76}{\vphantom{1.0}} & \stkcell{74.40}{0.99\,/\,0.99} & \stkcell{81.26}{\vphantom{1.0}} & \stkcell{97.92}{\vphantom{1.0}} & \stkcell{27.32}{0.63\,/\,0.63} & \stkcell{81.45}{\vphantom{1.0}} & \stkcell{96.32}{\vphantom{1.0}} & \stkcell{74.11\,{\scriptsize($\pm$0.63)}}{0.98\,/\,0.97} \\
 & k3m~\cite{DBLP:journals/amcs/SaeedTRA10} & \stkcell{90.31}{\textcolor[HTML]{1B7837}{\textbf{1.0}}\,/\,\textcolor[HTML]{1B7837}{\textbf{1.0}}} & \stkcell{93.97}{\vphantom{1.0}} & \stkcell{98.62}{\vphantom{1.0}} & \stkcell{48.10}{\textcolor[HTML]{1B7837}{\textbf{1.0}}\,/\,0.99} & \stkcell{68.15}{\vphantom{1.0}} & \stkcell{98.00}{\vphantom{1.0}} & \stkcell{74.97}{\textcolor[HTML]{1B7837}{\textbf{1.0}}\,/\,0.99} & \stkcell{76.85}{\vphantom{1.0}} & \stkcell{98.12}{\vphantom{1.0}} & \stkcell{64.74}{\textcolor[HTML]{1B7837}{\textbf{1.0}}\,/\,\textcolor[HTML]{1B7837}{\textbf{1.0}}} & \stkcell{75.89}{\vphantom{1.0}} & \stkcell{97.49}{\vphantom{1.0}} & \stkcell{79.88\,{\scriptsize($\pm$0.78)}}{\textcolor[HTML]{1B7837}{\textbf{1.0}}\,/\,0.99} \\
 & mahotas~\cite{DBLP:journals/corr/abs-1211-4907} & \stkcell{91.51}{\textcolor[HTML]{1B7837}{\textbf{1.0}}\,/\,0.99} & \stkcell{93.37}{\vphantom{1.0}} & \stkcell{98.93}{\vphantom{1.0}} & \stkcell{69.93}{\textcolor[HTML]{1B7837}{\textbf{1.0}}\,/\,0.97} & \stkcell{67.60}{\vphantom{1.0}} & \stkcell{98.27}{\vphantom{1.0}} & \stkcell{78.60}{\textcolor[HTML]{1B7837}{\textbf{1.0}}\,/\,0.99} & \stkcell{76.11}{\vphantom{1.0}} & \stkcell{98.58}{\vphantom{1.0}} & \stkcell{76.25}{\textcolor[HTML]{1B7837}{\textbf{1.0}}\,/\,0.99} & \stkcell{74.39}{\vphantom{1.0}} & \stkcell{98.58}{\vphantom{1.0}} & \stkcell{85.26\,{\scriptsize($\pm$0.41)}}{\textcolor[HTML]{1B7837}{\textbf{1.0}}\,/\,0.99} \\
 & stentiford~\cite{DBLP:journals/tsmc/StentifordM83} & \stkcell{89.98}{\textcolor[HTML]{1B7837}{\textbf{1.0}}\,/\,0.99} & \stkcell{94.20}{\vphantom{1.0}} & \stkcell{98.54}{\vphantom{1.0}} & \stkcell{32.11}{\textcolor[HTML]{1B7837}{\textbf{1.0}}\,/\,0.99} & \stkcell{68.15}{\vphantom{1.0}} & \stkcell{97.93}{\vphantom{1.0}} & \stkcell{67.95}{\textcolor[HTML]{1B7837}{\textbf{1.0}}\,/\,0.99} & \stkcell{76.59}{\vphantom{1.0}} & \stkcell{98.07}{\vphantom{1.0}} & \stkcell{37.40}{\textcolor[HTML]{1B7837}{\textbf{1.0}}\,/\,\textcolor[HTML]{1B7837}{\textbf{1.0}}} & \stkcell{76.16}{\vphantom{1.0}} & \stkcell{97.11}{\vphantom{1.0}} & \stkcell{74.59\,{\scriptsize($\pm$0.64)}}{\textcolor[HTML]{1B7837}{\textbf{1.0}}\,/\,0.99} \\
 & sitk~\cite{beare2018image} & \stkcell{83.65}{\textcolor[HTML]{1B7837}{\textbf{1.0}}\,/\,0.99} & \stkcell{96.06}{\vphantom{1.0}} & \stkcell{96.69}{\vphantom{1.0}} & \stkcell{29.08}{\textcolor[HTML]{1B7837}{\textbf{1.0}}\,/\,0.98} & \stkcell{70.69}{\vphantom{1.0}} & \stkcell{97.92}{\vphantom{1.0}} & \stkcell{63.27}{\textcolor[HTML]{1B7837}{\textbf{1.0}}\,/\,0.99} & \stkcell{79.17}{\vphantom{1.0}} & \stkcell{98.12}{\vphantom{1.0}} & \stkcell{29.79}{\textcolor[HTML]{1B7837}{\textbf{1.0}}\,/\,0.99} & \stkcell{78.55}{\vphantom{1.0}} & \stkcell{95.87}{\vphantom{1.0}} & \stkcell{68.98\,{\scriptsize($\pm$0.68)}}{\textcolor[HTML]{1B7837}{\textbf{1.0}}\,/\,0.99} \\
 & hilditch~\cite{hilditch1969linear} & \stkcell{30.08}{0.46\,/\,0.57} & \stkcell{95.80}{\vphantom{1.0}} & \stkcell{88.95}{\vphantom{1.0}} & \stkcell{9.48}{0.00\,/\,0.58} & \stkcell{71.60}{\vphantom{1.0}} & \stkcell{84.40}{\vphantom{1.0}} & \stkcell{21.90}{0.20\,/\,0.66} & \stkcell{79.64}{\vphantom{1.0}} & \stkcell{78.32}{\vphantom{1.0}} & \stkcell{13.32}{0.00\,/\,0.63} & \stkcell{79.43}{\vphantom{1.0}} & \stkcell{79.82}{\vphantom{1.0}} & \stkcell{24.66\,{\scriptsize($\pm$1.01)}}{0.32\,/\,0.59} \\
\midrule
\multirow{2}{*}{\raisebox{0pt}[0pt][0pt]{\rotatebox[origin=c]{90}{\scriptsize\itshape Learned}}} & skelite~\cite{DBLP:conf/ipmi/VargasMPNA25} & \stkcell{41.29}{0.32\,/\,\textcolor[HTML]{1B7837}{\textbf{1.0}}} & \stkcell{44.09}{\vphantom{1.0}} & \stkcell{99.93}{\vphantom{1.0}} & \stkcell{54.86}{0.00\,/\,0.98} & \stkcell{63.61}{\vphantom{1.0}} & \stkcell{99.80}{\vphantom{1.0}} & \stkcell{72.27}{0.53\,/\,0.99} & \stkcell{70.78}{\vphantom{1.0}} & \stkcell{99.83}{\vphantom{1.0}} & \stkcell{70.95}{0.00\,/\,0.97} & \stkcell{70.52}{\vphantom{1.0}} & \stkcell{99.87}{\vphantom{1.0}} & \stkcell{49.88\,{\scriptsize($\pm$1.02)}}{0.28\,/\,0.99} \\
 & unet~\cite{DBLP:conf/iccvw/Nguyen21} & \stkcell{0.04}{0.00\,/\,0.01} & \stkcell{88.63}{\vphantom{1.0}} & \stkcell{42.88}{\vphantom{1.0}} & \stkcell{0.01}{0.00\,/\,0.03} & \stkcell{49.27}{\vphantom{1.0}} & \stkcell{86.20}{\vphantom{1.0}} & \stkcell{5.50}{0.04\,/\,0.20} & \stkcell{50.78}{\vphantom{1.0}} & \stkcell{85.69}{\vphantom{1.0}} & \stkcell{1.49}{0.00\,/\,0.05} & \stkcell{49.76}{\vphantom{1.0}} & \stkcell{83.35}{\vphantom{1.0}} & \stkcell{0.98\,{\scriptsize($\pm$0.09)}}{0.00\,/\,0.05} \\
\bottomrule
\end{tabular}
 \end{table*}

\subsubsection{Skeleton Reconstruction Quality Score (SkelReconQS)}
\label{subsubsec:srqs_metric}

SkelReconQS rewards a compact skeleton only when it reconstructs the input
faithfully. For an object mask $O$ and skeleton $S$,
\begin{equation}
\mathrm{SkelReconQS} = T_{\mathrm{recon}}\,\mathrm{Conn}\,
C^{1/4}\bigl(F\,G_{\mathrm{avg}}\,G_{\mathrm{om}}\bigr)^{1/4},
\label{eq:skelqs}
\end{equation}
where $T_{\mathrm{recon}}$, $\mathrm{Conn}$ are undamped structural factors,
graded in $[0,1]$ rather than binary, and $C$, $F$, $G_{\mathrm{avg}}$,
$G_{\mathrm{om}}$ the damped compaction and reconstruction terms. The
diagnostics $T_{\mathrm{skel}}$, $\mathrm{Dice}_{\mathrm{sp}}$ are reported
separately and do not enter it. All terms are read per image.

\emph{Common decoder.} Label $O$ into $N$ objects $O_i$ under the evaluation
connectivity ($8$-connected throughout the main benchmark) and let $S_i$ be the skeleton pixels on $O_i$; mass off every object is discarded. Let $D_i(x)$ be the exact Euclidean distance from $x$ to the nearest background-pixel centre of $O_i$'s isolated crop, its bounding box padded by one background pixel so that no neighbour contributes. Then
\begin{equation}
r_i(p) = \bigl\lceil D_i(p)\bigr\rceil,\quad
R_i = \bigcup_{p\in S_i}\overline{B}\bigl(p,r_i(p)\bigr),\quad
R = \bigcup_i R_i,
\label{eq:srqs_decode}
\end{equation}
with $\overline{B}$ the closed disc on the image lattice
$\Lambda\subset\mathbb{Z}^2$, so every disc is truncated at the frame, and
$R_i^{\mathrm{clip}} = R_i\cap O_i$. Every method is decoded by
\eqref{eq:srqs_decode}, so no SpiderSkel Stage-3 radius $r_{\ast}$ enters
SkelReconQS and the sub-pixel offsets reach only $\mathrm{Dice}_{\mathrm{sp}}$.
The reconstruction terms score the raw union $R$, keeping spill and omission visible, while the topology certificate reads $R_i^{\mathrm{clip}}$, hence shape-constrained.

\emph{Damped terms.} Let $\partial_n X$ collect the pixels of $X$ having an
$n$-neighbour outside $X$, the canvas exterior counting as outside, and let
$d(x,Y)$ be the Euclidean distance from a point $x$ to the nearest pixel of
$Y$. With
$d_{\mathrm{om}} = \max_{q\in O} d(q,R)$ the worst-case omission depth and
$d_{\mathrm{ref}} = |O|/|\partial_8 O|$ the characteristic half-thickness
that normalises both distance penalties,
\begin{equation}
\begin{aligned}
C &= \frac{|O|-|S|+1}{|O|+1},\quad
\mathrm{Reduction} = \frac{|O|-|S|}{|O|},\\
F &= \mathrm{Dice}(O,R) = \frac{2\lvert O\cap R\rvert}{\lvert O\rvert+\lvert R\rvert},\\
\overline{d}(X,Y) &= \mathop{\mathrm{mean}}\limits_{p\in\partial_4 X} d(p,\partial_4 Y),\;
\mathrm{ASSD} = \tfrac{\overline{d}(O,R)+\overline{d}(R,O)}{2},\\
G_{\mathrm{avg}} &= e^{-\mathrm{ASSD}/d_{\mathrm{ref}}},\quad
G_{\mathrm{om}} = e^{-\max(0,\,d_{\mathrm{om}}-\sqrt{2})/d_{\mathrm{ref}}}.
\end{aligned}
\end{equation}
The boundary subscripts differ as implemented: $\mathrm{ASSD}$ is read on
$\partial_4$, $d_{\mathrm{ref}}$ on $\partial_8$; the $\sqrt{2}$ offset in
$G_{\mathrm{om}}$ is one diagonal pixel of tolerance. Here $|O|$, $|S|$ are
whole-image counts pooled over objects; only $C$ enters
\eqref{eq:skelqs}; $\mathrm{Reduction}$ is the separate column of
Table~\ref{tab:category_avg_skelqs_em}.

\emph{Structural factors.} Object $i$ passes when the clipped reconstruction
remains connected and preserves the object's holes one-to-one. Precisely, let
$\mathcal{H}(X)$ be the bounded background components of $X$ under the dual
connectivity ($4$-connected here), read on each object's crop padded by one
background pixel. Since $R_i^{\mathrm{clip}}\subseteq O_i$, each hole of $O_i$
lies in exactly one background component of $R_i^{\mathrm{clip}}$; let $\iota_i$
send it there when that component is bounded, and leave $\iota_i$ undefined when
the hole has opened to the exterior. Writing $n_{\mathrm{cc}}^{(8)}(\cdot)$ for
the number of $8$-connected components and $k_i = n_{\mathrm{cc}}^{(8)}(S_i)$,
\begin{equation}
\begin{aligned}
t_i &= \mathbb{1}\!\left[
\begin{aligned}
&n_{\mathrm{cc}}^{(8)}\bigl(R_i^{\mathrm{clip}}\bigr)=1,\quad
\operatorname{dom}(\iota_i)=\mathcal{H}(O_i),\\
&\operatorname{im}(\iota_i)=\mathcal{H}\bigl(R_i^{\mathrm{clip}}\bigr),\\
&\forall h,h'\in\mathcal{H}(O_i)\colon\
\iota_i(h)=\iota_i(h')\Rightarrow h=h'
\end{aligned}
\right],\\
T_{\mathrm{recon}} &= \Bigl(\tfrac{1}{N}\textstyle\sum_i t_i\Bigr)\,
\mathbb{1}\bigl[\forall i\colon R_i^{\mathrm{clip}}\neq\emptyset\bigr],\quad
\mathrm{Conn} = \tfrac{1}{N}\textstyle\sum_i \tfrac{1}{k_i},
\end{aligned}
\end{equation}
with $1/k_i = 0$ by convention for an object that received no skeleton. The
missing-object indicator is the only hard gate.

\emph{Diagnostics.} $T_{\mathrm{skel}}$ certifies the skeleton against the shape
rather than the reconstruction: per image, under the $(8,4)$
foreground/background pairing on the whole canvas, the $8$-connected skeleton
components are in bijection with the object components and each matched pair
agrees in hole count; off-object mass, discarded by the decoder, does not affect this certificate. The reported figure is the fraction of images passing. Strict sub-pixel
reconstruction Dice
decodes at the method's own sub-pixel centres $\widetilde{S}_i$ under an open
disc,
\begin{equation}
\begin{aligned}
R_{\mathrm{sp}} &= \bigcup_i \bigcup_{c\in\widetilde{S}_i}
  \bigl\{q\in\Lambda : \lVert q-c\rVert^2 < D_i(c)^2\bigr\},\\
\mathrm{Dice}_{\mathrm{sp}} &= \mathrm{Dice}(O,R_{\mathrm{sp}}),
\end{aligned}
\label{eq:dice_sp}
\end{equation}
with the radius read from the field $D_i$ of \eqref{eq:srqs_decode},
now at the sub-pixel centre rather than rounded up at the lattice point, so no
Stage-3 radius is inherited; methods reporting no sub-pixel estimate decode at
integer centres under the same rule.

\subsubsection{Evaluation Protocol}

Headline results use whole-image execution, with object-aware decoding and
structural evaluation as defined above.
Componentwise execution, one object at a time in its own crop, is the protocol
used for the studies Supplementary
Appendix~\ref{app:evaluation_protocol} lists.
Supplementary Appendix~\ref{app:evaluation_protocol} defines both protocols,
states the frame-handling rule shared by every method, and summarises the
cross-protocol comparison.

\subsubsection{Threshold Selection}

Our method's angular thresholds $(\phi_l, \phi_h)$ on the selected span
$\widehat S_P$ (\cref{eq:phi_thresholds} of
Section~\ref{sec:Proposed Method}) trade
compactness against reconstruction fidelity in opposite directions and admit
no closed-form optimum, so we sweep a $9\times9=81$-point grid, $\phi_l$ from
$0$ to $\pi/2$ and $\phi_h$ from $\pi/2$ to $\pi$ in steps of $\pi/16$. The
operating point $(0,13\pi/16)$ is the SkelReconQS argmax of the $81$ cells
over the evaluation corpus itself
and is fixed for every downstream evaluation, except where a sensitivity or ablation study varies it; at $\phi_l=0$ the
lower cutoff prunes only zero spans, leaving one cutoff plus the support
test in effect.
The remaining non-$\phi$ hyperparameters stay at their defaults (representative-retention cap $N_{\max}=20$); Supplementary Appendix~\ref{app:threshold_search}
records the sweep, the leave-one-category-out bound, and a one-at-a-time
sensitivity audit.

\subsubsection{Statistical Reporting}

Every column of Table~\ref{tab:category_avg_skelqs_em} is macro-averaged over
datasets. The Overall column's
$(\pm h)$ in
Table~\ref{tab:category_avg_skelqs_em} is a 95\% percentile-bootstrap
confidence-interval half-width, images resampled within their dataset
($B{=}10{,}000$ paired replicates), conditional on the fixed dataset collection
rather than between-dataset sampling uncertainty. Stars are Holm-corrected
two-sided paired bootstrap tests on the same estimand, reported for the Overall
column against the baseline scoring highest there, \texttt{medial\_axis}. Every
method enters the Holm family; stars are displayed only for SpiderSkel rows.
Bootstrap resolution
floors $p$ at $2/B$ and Holm over the twenty-five
methods multiplies the smallest by twenty-five, so $p_{\mathrm{Holm}}$ floors at
$0.005$ and no finer tier is reported. Displayed certificate fractions below one floor to two decimals; a
near-miss never displays as perfect.

\FloatBarrier

\subsection{Results}

The results are read against the contributions: the overall comparison and
its certificates evaluate the complete method and support contribution (v); the component substitutions
isolate the mechanisms of (i)--(iii); and the robustness and
incremental-update subsections stress the same operating point under
degradation and editing, completing (v) and testing (iv).

\subsubsection{Overall Comparison}
\label{subsubsec:overall_comparison}

SpiderSkel-N8 and SpiderSkel-DualC attain an Overall SkelReconQS of
$91.09\%$, with SpiderSkel-Scaffold ($90.94\%$) close behind,
all three significantly above the strongest baseline, \texttt{medial\_axis}
($86.62\%$), under the Holm-corrected paired bootstrap ($p{<}0.01$).

The three SpiderSkel variants are the only evaluated methods attaining exact
skeleton topology, full connectivity and exact per-object reconstruction
topology ($T_{\mathrm{skel}}=\mathrm{Conn}=T_{\mathrm{recon}}=1$) in every
category aggregate: the closest baseline, hjs,
attains all three everywhere except document binarisation, where it
misses $T_{\mathrm{recon}}$ on one image of the $11{,}323$. Retuned to
$\gamma{=}5.0$ on the evaluation corpus (Supplementary
Appendix~\ref{app:hjs_gamma}), hjs scores $0.909$ SkelReconQS and $0.901$
Reduction under that corpus-level protocol against $0.911$ and $0.906$ for
SpiderSkel-N8, and $27$ of its images keep $G_{\mathrm{om}}{<}0.9$ against
none: aggregate overlap and Reduction can understate deep, local omission.

Each of the three is Pareto-optimal over all twenty-two baselines in
(Dice$_{\mathrm{sp}}$, Reduction Rate) in all four categories
(per-category columns, Table~\ref{tab:category_avg_skelqs_em}; per-category
reading, Supplementary Appendix~\ref{app:overall_full}), scoped to the
strict sub-pixel decode of \eqref{eq:dice_sp}: no baseline
reconstructs at strict tolerance as accurately at comparable reduction, and the tolerance-matching
learned dense predictor (skelite) reaches only
$44$--$71\%$ Reduction Rate against our $76$--$96\%$.

The certificates are an
empirical record of no observed failure, not a homotopy-equivalence claim.

Over all $11{,}323$ images and their $31{,}209$ objects, the three SpiderSkel variants violate neither certificate anywhere;
head-to-head on every image against four classical baselines, no
SpiderSkel variant loses a certificate that one of them holds, and where it
loses the composite to a baseline it trails that baseline by at most $0.5$~pp
of strict sub-pixel Dice (Supplementary Appendix~\ref{app:failure_analysis}). The surviving trade-off is density on
shapes two or three pixels wide, which the optional connectivity-aware
reduction trims.

\subsubsection{Runtime and Computational Cost}
\label{subsec:runtime}

\begin{figure}[!t]
  \centering
  \includegraphics[width=\columnwidth]{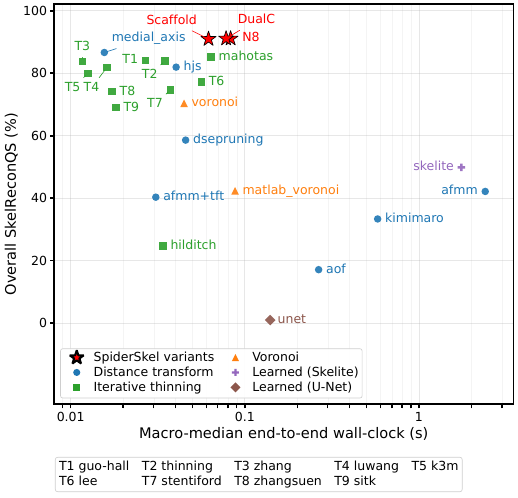}
  \caption{Macro-median end-to-end wall-clock (log scale, $x$-axis) against the
  Overall SkelReconQS of Table~\ref{tab:category_avg_skelqs_em} ($y$-axis). The
  three SpiderSkel variants nearly coincide at this scale; the dense
  iterative-thinning cluster is keyed T1--T9 below the plot. The two learned
methods appear as their released models.}
  \label{fig:runtime_vs_skelqs}
\end{figure}

Each SpiderSkel variant is Pareto-optimal in wall-clock time against Overall
SkelReconQS over all twenty-two baselines (Fig.~\ref{fig:runtime_vs_skelqs}).
The variants take roughly $4$--$7\times$ as long as the fastest classical
baselines ($0.062$--$0.083$~s against $0.012$--$0.016$~s), but
every method faster than a given variant scores at least $4.3$ points below
it in Overall SkelReconQS (binding case \texttt{medial\_axis} against
SpiderSkel-Scaffold; Table~\ref{tab:category_avg_skelqs_em}); the multi-tangent
interpolation dominates the shared core's cost.

Asymptotic cost differs by family. Iterative thinning is $\mathcal{O}(N\cdot\text{iters})$. The distance-based baselines share a linear exact Euclidean distance transform, so their post-processing separates them: an $\mathcal{O}(N\log N)$ ordering step for \texttt{medial\_axis} and \texttt{hjs}, and for \texttt{afmm} a per-pixel junction span of $\mathcal{O}(N\!\cdot\!J)$ in the junction count $J$. The Voronoi pair is $\mathcal{O}(N{+}V\log V)$ in its boundary-sample count $V$, and our multi-tangent core is $\mathcal{O}(N{+}T)$ in the tied-site mass $T$ (Supplementary Appendix~\ref{app:envelope}).

Protocol: timings cover the same methods on a stratified 16-image
subset (four per category); every cell is the median of
three independent occasions, each on an exclusive node with one
strictly serial thread (maximum inter-occasion spread
$4.7\%$, $2.2\%$ for the three proposed methods), macro-medianed
within and then across categories for robustness to image-size
outliers (summary: Supplementary Appendix~\ref{app:runtime}).

\subsubsection{Score Composition}
\label{subsubsec:score_composition}
The composite is multiplicative, so wherever the score is nonzero
$\log\mathrm{SkelReconQS}$ is the weighted sum of the six log-factors, weight
one on the structural factors and $1/4$ on the damped terms, and each factor's
contribution to the between-method separation decomposes exactly. Exactly-zero scores, whose logarithm is undefined ($6.3\%$ of method-image
cells, none from a SpiderSkel variant), are counted separately, not imputed.

Which term discriminates is
category-dependent: connectivity on solid shapes, and on the three
thin-structure categories the omission factor, the only term penalising
the maximum omission depth. The ranking depends materially on that term alone: holding
the omission factor at $1$ drops the Spearman correlation with the benchmark
ranking to $0.907$ and moves hjs to first, whereas every other
single-term drop leaves the three variants as the top three. Near-saturated terms bind: under-compaction can raise omission, which
compaction penalises.

Scaling each of the six exponents independently by $1/2$, $1$ or $2$ gives
$3^{6}=729$ positive reweightings, in all of which the three SpiderSkel
variants remain the top-three set (Supplementary Appendix~\ref{app:score_composition}); only removing the omission
term, as above, changes the ranking, and the per-category
(Dice$_{\mathrm{sp}}$, Reduction Rate) Pareto claim is independent of the
composite.

\subsubsection{Ablation}
\label{subsubsec:component_substitution}

\begin{table*}[!t]
\centering
\footnotesize
\setlength{\tabcolsep}{3.2pt}
\renewcommand{\arraystretch}{1.05}
\caption{Component substitutions: each advocated shared-core component replaced
by a simpler or denser alternative, all else fixed, for the three
SpiderSkel variants, componentwise on all 11{,}323 benchmark images.
Unmodified runs reproduce the main benchmark's skeletons image for
image. ``Affected diagnostics'' lists the
exact-topology fraction $T_{\mathrm{skel}}$ (N8/DualC/Scaffold),
skeleton density (Reduction), and strict sub-pixel reconstruction
Dice (Dice$_{\mathrm{sp}}$), the latter two worst-of-three. The Scaffold variant has no
offset-safety dial: its offset construction is the anchor architecture
itself}
\label{tab:component_substitution}
\begin{tabular}{@{}ll ccc l@{}}
\toprule
 & & \multicolumn{3}{c}{$\Delta$SkelReconQS (pp, Overall)} & \\
\cmidrule(lr){3-5}
Component & Substitution & SpiderSkel-N8 & SpiderSkel-DualC & SpiderSkel-Scaffold & Affected diagnostics \\
\midrule
Geometry-adaptive neighbourhood & removed (center only) & $-82.55$ & $-82.76$ & $-29.11$ & $T_{\mathrm{skel}}$ 0.888/0.705/1; Dice$_{\mathrm{sp}}$ $-68.4$\,pp \\
 & main axis only & $-24.67$ & $-25.34$ & $-0.04$ & $T_{\mathrm{skel}}$ 0.927/0.867/1; Dice$_{\mathrm{sp}}$ $-2.0$\,pp \\
 & fixed 4-neighbourhood & $-0.90$ & $-0.90$ & $-0.04$ & $T_{\mathrm{skel}}$ 0.990/0.990/1 \\
 & all 8 neighbours & $+0.01$ & $+0.01$ & $+0.10$ & reduction $-0.2$\,pp \\
\addlinespace[2pt]
Multi-label tangent set & single label & $-0.04$ & $-0.04$ & $+0.08$ & -- \\
 & capacity 20$\to$2 & $+0.01$ & $+0.01$ & $0.00$ & -- \\
\addlinespace[2pt]
Support sampling ($N_{\mathrm{s}}{=}15$) & $N_{\mathrm{s}}{=}1$ & $-16.51$ & $-21.30$ & $-0.04$ & $T_{\mathrm{skel}}$ 1/0.961/1; Dice$_{\mathrm{sp}}$ $-1.2$\,pp \\
 & $N_{\mathrm{s}}{=}3$ & $-0.78$ & $-1.35$ & $0.00$ & $T_{\mathrm{skel}}$ 1/0.999/1 \\
 & $N_{\mathrm{s}}{=}7$ & $-0.08$ & $-0.16$ & $+0.01$ & -- \\
 & $N_{\mathrm{s}}{=}31$ & $-0.02$ & $-0.02$ & $-0.01$ & -- \\
\addlinespace[2pt]
Span ($\varphi$) filter & disabled & $-0.08$ & $-0.08$ & $0.00$ & $T_{\mathrm{skel}}$ 0.999/0.999/1 \\
\addlinespace[2pt]
Connect-aware offset & no connectivity guard & $0.00$ & $-0.07$ & -- & $T_{\mathrm{skel}}$ 1/0.999/-- \\
\addlinespace[2pt]
Sub-pixel offsets & integer decode & $0.00$ & $0.00$ & $0.00$ & Dice$_{\mathrm{sp}}$ $-1.0$\,pp \\
\bottomrule
\end{tabular}
 \end{table*}

The substitutions of Table~\ref{tab:component_substitution}
separate the components into two classes. The first carries the score:
removing the geometry-adaptive neighbourhood or collapsing support sampling
to $N_{\mathrm{s}}{=}1$ costs up
to $83$~pp of SkelReconQS and reduces the exact-topology fraction in three of those six
cells, while the sampling
sweep brackets the default $N_{\mathrm{s}}{=}15$ as an elbow.
SpiderSkel-Scaffold is the exception: its anchored erosion absorbs the
sampling collapse almost entirely ($-0.04$~pp), a property of the
variant's architecture, not the substitution. The neighbourhood rows
quantify candidate burden: the \emph{all} substitution marks all eight
stencil neighbours where the adaptive rule marks at most two per reference
site, and buys at most $+0.10$~pp, so the certificates hold on far fewer
candidates.

The second class carries the certificates rather than the score: disabling
the span filter or composing the connect-aware offset without the connectivity guard
moves SkelReconQS by at most $0.08$~pp yet violates
exactness on a few images, and discarding the sub-pixel offsets is
invisible to the integer-decoded score while forfeiting strict sub-pixel
representation. Densifying substitutions can edge the
default by up to $0.10$~pp, the reconstruction terms rewarding
coverage slightly ahead of the reduction penalty, so
SkelReconQS is read jointly with the certificate and reduction columns.

The tie-aware working-set rows probe contribution~(i) directly: collapsing
the multi-label tangent set to a single label moves Overall by at most
$0.08$~pp with no certificate broken, largely because downstream
interpolation and aggregation recover much of the discarded evidence, and
the cap sweep $N_{\max}\in\{2,5,10,20\}$ moves the displayed macro score by at
most $0.01$~pp while changing individual outputs on $69$ of $135$ audited images
(Supplementary Appendix~\ref{app:threshold_search}); the explicit tie record is
retained for completeness and determinism.

Together these substitutions ground contributions (ii) and (iii): the
interpolation of (ii) carries the score, the span filter and connect-aware
offset of (iii) carry the certificates, and the sub-pixel offsets carry
strict sub-pixel fidelity, distinctions the composite alone does not
capture.

\subsection{Robustness Analysis}
\label{subsec:robustness}

\begin{figure}[!htbp]
  \centering
  \includegraphics[width=\linewidth]{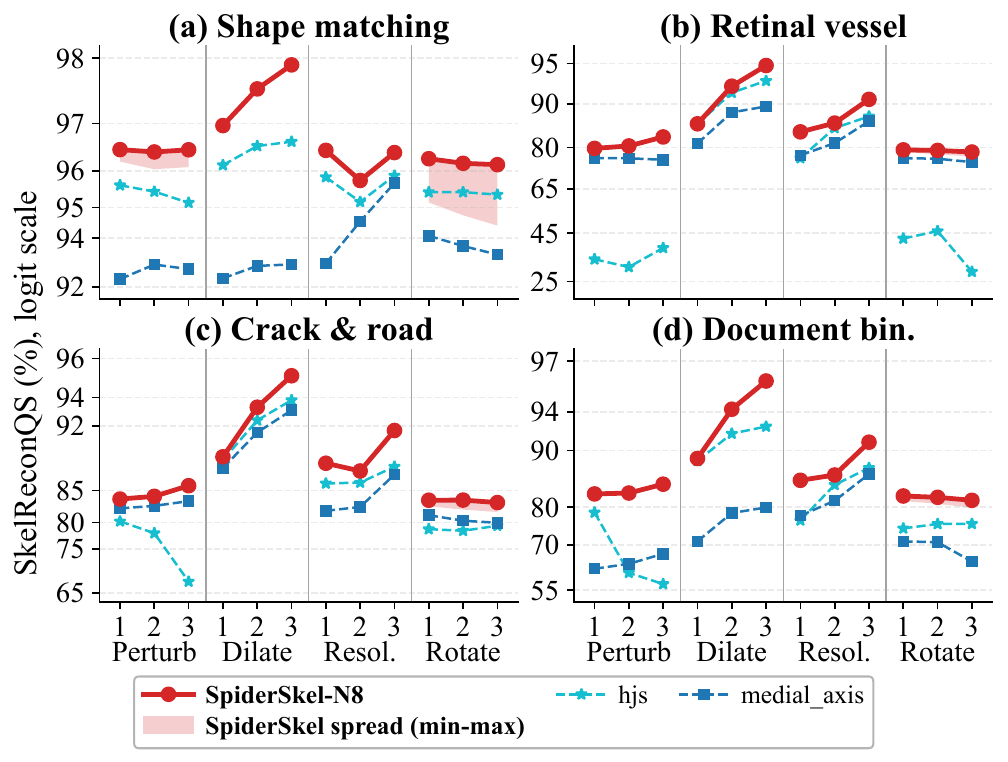}
  \caption{Absolute SkelReconQS (\%) under the twelve-condition robustness
  protocol on the paired 2{,}715-image subset, per benchmark category. Each
degradation axis has three severities: Perturb (amplitude
  $a=0.5,1.0,2.0$), Dilate (radius $k=1,3,5$), Resol.\ (factor
  $2\times,4\times,8\times$) and Rotate ($15^\circ,30^\circ,45^\circ$);
  lines are not joined across axis groups. The $y$ axis is the logit
  $\log\!\big(s/(100-s)\big)$ of the score $s$, ticks labelled in score
  units; panels use independent $y$ limits, and vertical distances compare
  only within a panel. Red line:
  SpiderSkel-N8; band: the three variants' min--max envelope.}
  \label{fig:robustness_per_category}
\end{figure}

Table~\ref{tab:robustness_skelqs} pools image-level means,
Fig.~\ref{fig:robustness_per_category} shows dataset macros, and quoted
minima are per category.

We stress-test along four complementary distortion axes: \emph{boundary
perturbation} (high-frequency contour noise), \emph{dilation} (low-frequency
foreground thickening), \emph{resolution degradation} (aggressive
downsample$\,\to\,$upsample), and \emph{rotation} (resampling onto a rotated
pixel lattice), each at three severities, on a nineteen-dataset
subset fixed before analysis, stratified by category (Supplementary
Appendix~\ref{app:robustness_full_tables}).

Scores are computed against the
input the method actually received,
measuring robustness of extraction rather than recovery of removed
information: a degraded input can be simpler to reconstruct, so scores may
rise with severity and only within-condition comparisons are meaningful. SpiderSkel holds one operating point
across datasets and severities; baselines keep their configurations.

\begin{table}[!htb]
\centering
\scriptsize
\setlength{\tabcolsep}{2pt}
\renewcommand{\arraystretch}{1.05}
\caption{SkelReconQS (\%) under the four degradation axes
(\emph{Perturb.}\ = boundary perturbation, \emph{Dilat.}, \emph{Resol.},
\emph{Rot.}) at three severities each, on the deterministic 2{,}715-image
evaluation subset. Bold shaded rows are the SpiderSkel variants
\mbox{N8}, DualC and Scaffold.
Each cell is the pooled mean of the method's per-image scores on the
degraded input; the dataset-macro view is
Fig.~\ref{fig:robustness_per_category}}
\label{tab:robustness_skelqs}
\begin{tabular}{@{}l c c c c c c c c c c c c @{}}
\toprule
 & \multicolumn{3}{c}{\textit{Perturb.}} & \multicolumn{3}{c}{\textit{Dilat.}} & \multicolumn{3}{c}{\textit{Resol.}} & \multicolumn{3}{c}{\textit{Rot.}} \\
\cmidrule(lr){2-4} \cmidrule(lr){5-7} \cmidrule(lr){8-10} \cmidrule(lr){11-13}
Method & Mil. & Mod. & Sev. & Mil. & Mod. & Sev. & Mil. & Mod. & Sev. & Mil. & Mod. & Sev. \\
\midrule
\cellcolor{black!8}\textbf{N8} & \cellcolor{black!8}94.2 & \cellcolor{black!8}94.2 & \cellcolor{black!8}94.5 & \cellcolor{black!8}95.5 & \cellcolor{black!8}96.9 & \cellcolor{black!8}97.5 & \cellcolor{black!8}94.9 & \cellcolor{black!8}94.3 & \cellcolor{black!8}95.6 & \cellcolor{black!8}93.9 & \cellcolor{black!8}93.8 & \cellcolor{black!8}93.7 \\
\cellcolor{black!8}\textbf{DualC} & \cellcolor{black!8}94.2 & \cellcolor{black!8}94.2 & \cellcolor{black!8}94.5 & \cellcolor{black!8}95.5 & \cellcolor{black!8}96.9 & \cellcolor{black!8}97.5 & \cellcolor{black!8}94.9 & \cellcolor{black!8}94.3 & \cellcolor{black!8}95.6 & \cellcolor{black!8}93.9 & \cellcolor{black!8}93.8 & \cellcolor{black!8}93.7 \\
\cellcolor{black!8}\textbf{Scaffold} & \cellcolor{black!8}94.0 & \cellcolor{black!8}93.8 & \cellcolor{black!8}94.2 & \cellcolor{black!8}95.6 & \cellcolor{black!8}96.9 & \cellcolor{black!8}97.5 & \cellcolor{black!8}94.8 & \cellcolor{black!8}94.3 & \cellcolor{black!8}95.6 & \cellcolor{black!8}92.9 & \cellcolor{black!8}92.5 & \cellcolor{black!8}92.1 \\
\midrule
guo-hall & 87.1 & 87.3 & 87.1 & 86.4 & 85.7 & 85.1 & 87.0 & 84.3 & 81.6 & 88.0 & 87.5 & 86.5 \\
hjs & 91.2 & 90.0 & 89.1 & 94.8 & 95.8 & 96.1 & 93.5 & 93.6 & 94.6 & 91.1 & 91.3 & 90.8 \\
mahotas & 89.1 & 88.8 & 88.7 & 89.0 & 90.2 & 91.0 & 89.3 & 89.9 & 90.9 & 88.4 & 87.9 & 87.0 \\
medial\_axis & 90.0 & 90.5 & 90.5 & 91.0 & 92.4 & 92.7 & 91.0 & 92.5 & 94.3 & 91.4 & 90.9 & 90.2 \\
thinning & 87.0 & 87.2 & 87.1 & 86.3 & 85.7 & 85.0 & 86.8 & 84.2 & 81.5 & 87.9 & 87.5 & 86.5 \\
zhang & 87.1 & 87.0 & 87.0 & 87.2 & 88.2 & 88.9 & 88.9 & 87.0 & 85.0 & 87.6 & 88.1 & 84.2 \\
\bottomrule
\end{tabular}
 \end{table}

Under the fixed operating point, the variants post the highest
absolute score in the nine-method roster (the variants plus the six
strongest baselines by Overall score) in all twelve
conditions (Table~\ref{tab:robustness_skelqs}), the robustness half of
contribution~(v). Those pooled means hide a per-category
collapse that
Fig.~\ref{fig:robustness_per_category} resolves: on retinal vessels the
flux-based hjs breaks down under boundary perturbation and rotation,
trailing the best variant by $45$--$50$~pp at worst severity; no variant
falls below $77.1\%$ in any condition.

Dilation thickens objects and raises scores; elsewhere they move little:
pooled over the $2{,}715$-image subset, no per-condition mean change from the
corresponding clean-input score falls below $-1.9$ or rises above $+3.5$~pp. On shape
matching under rotation, though, hjs scores above the weakest SpiderSkel
variant at all three severities
(Fig.~\ref{fig:robustness_per_category}, rotation group).

\subsection{Incremental Update}
\label{sec:incremental}
The incremental engines are required to reproduce exactly the skeleton of a complete SpiderSkel run on the edited mask. Across $23{,}040$ verified
edit steps ($240$ images${}\times{}2$ workloads${}\times{}16$
steps${}\times{}3$ variants) every incremental
output equals the skeleton of a complete recomputation of the same mask, and the
incremental engine, the retention-only engine and the complete
recomputation agree on every sequence's final skeleton, establishing the
equivalence claim of contribution~(iv).

The multi-tangent interpolation
dominates the runtime measured above, so we maintain the
variants under local edits rather than recompute them: the engine holds one entry per object, its crop and persisted
candidate field, untouched objects keep their skeletons unchanged, and a touched object
re-derives through the shared core, falling back to a per-object recomputation
when the field cannot certify the edit. That
componentwise decomposition is a deployment choice
(Supplementary Appendix~\ref{app:evaluation_protocol}).

Incremental
distance-field and free-space-skeleton maintenance is established prior
art, the closest being \textit{Lau et al.}'s incrementally maintained
generalized Voronoi diagrams~\cite{DBLP:conf/iros/LauSB10}; none meets the
exactness requirement above.

The timing protocol is paired: every cell of every workload pools the same
$6{,}000$ per-step pairs ($240$ images $\times$ $25$ steps), an incremental
update and a complete
recomputation of the same post-edit mask, run back-to-back in seeded order
(C++ implementation, one exclusive node, single-threaded, compiled defaults). Confidence intervals are percentile bootstraps resampled at the image
level, since the $25$ steps on one image share its layout and size.
Pooled per-step medians for the paired complete recomputation are $71$\,ms
for N8 and $56$\,ms for Scaffold under both workloads, and $98$\,ms
(foreground) to $99$\,ms (canvas) for DualC; \emph{Marginal} divides
median-of-medians recomputation by median-of-medians update time, not
per-step ratios; the \emph{Text stratum} restricts the paired ratio to the $1{,}200$ document-text pairs, where an
edit touches few of the many small glyphs, so retention dominates even
foreground edits.

Paired timing over this fixed stratified sample
(Table~\ref{tab:incsafe_paired}) gives $1.6$--$2.6\times$ faster foreground
edits and $21$--$33.0\times$ faster canvas edits: the persisted field
carries foreground edits, componentwise retention carries canvas edits, and
the worst case approaches one complete recomputation of the object. The procedure, edit generator, exactness checks and speedup
decomposition are in Supplementary
Appendix~\ref{app:incremental_algorithm}.

\begin{table}[!htb]
\centering
\footnotesize
\setlength{\tabcolsep}{4pt}
\caption{Paired incremental timing of the no-reduce variants over
  the fixed stratified sample, for foreground and canvas edit workloads.
  Paired: median speedup of the incremental update over the complete
  recomputation ($\times$, with $95\%$ confidence interval); Update: the
  incremental update's pooled per-step median wall-clock (ms); the
  Text-stratum and Marginal columns are defined in the text}
\label{tab:incsafe_paired}
\begin{tabular}{@{}l ccc ccc@{}}
\toprule
 & \multicolumn{3}{c}{Foreground edits} & \multicolumn{3}{c}{Canvas edits} \\
\cmidrule(lr){2-4} \cmidrule(lr){5-7}
Method & \stkcell{Paired}{[CI$_{95}$]} & \stkcell{Update}{[ms]} & \stkcell{Text}{stratum} & \stkcell{Paired}{[CI$_{95}$]} & \stkcell{Update}{[ms]} & Marginal \\
\midrule
N8 & \stkcell{1.55$\times$}{[1.52,1.58]} & 18.1 & 17.37$\times$ & \stkcell{26.35$\times$}{[25.31,27.25]} & 4.4 & 19.71$\times$ \\
DualC & \stkcell{2.61$\times$}{[2.42,2.79]} & 16.4 & 29.06$\times$ & \stkcell{32.96$\times$}{[31.66,34.01]} & 4.4 & 27.46$\times$ \\
Scaffold & \stkcell{1.67$\times$}{[1.63,1.71]} & 15.4 & 15.95$\times$ & \stkcell{21.14$\times$}{[20.51,21.94]} & 4.3 & 16.16$\times$ \\
\bottomrule
\end{tabular}
 \end{table}

\subsection{Application: Skeleton-to-Graph Analysis}
\label{sec:graph}
A one-pixel skeleton converts
deterministically to a graph: endpoints and junctions become nodes, runs between them edges, and
detected cycles closed loops, carrying
per-edge local radius profiles, junction angles and, for trees,
Horton--Strahler order~\cite{horton1945erosional,strahler1957quantitative}. This is a scoped feasibility demonstration; procedure, examples
and cross-language parity await publication.
 
\section{Conclusion}
We presented a 2D skeletonization pipeline that combines exact Euclidean
distance computation with a bounded, tie-aware nearest-site working set,
multi-tangent subpixel interpolation, angular significance filtering, and
connectivity-aware refinement. The contribution is this integration and its
operating-point analysis; the exact feature-transform set it builds on is
established prior art.

The evidence traces the contributions: the substitutions show the
interpolation carrying the score and the filtering stages the certificates
((ii), (iii)), the tie-aware working set retaining
complete, deterministic evidence (i); the incremental engines
match recomputation exactly on every verified edit (iv); and at one fixed
operating point the three variants alone hold the structural certificates
everywhere, remain Pareto-optimal per category in (Dice$_{\mathrm{sp}}$,
Reduction Rate), and lead the six
strongest baselines in every robustness condition (v).

\subsection{Limitations}
First, the connection and cleaning stages of SpiderSkel-N8 and
SpiderSkel-DualC are deterministic heuristics: their rules protect
8-connected foreground structure under local tests. SpiderSkel-Scaffold runs
neither, eroding onto anchors by $(8,4)$ simple-point deletion alone, so
component and cycle counts hold by construction. A digital-homotopy proof
for the complete pipelines remains open; failure counts accompany the
certificates.

Second, disc-based reconstruction rewards short medial responses to
boundary roughness: thin road masks can retain an uncovered terminal stub, and near-resolution silhouettes can be denser than one-pixel
thinning; the score penalises spill and omission on the raw decode, the
topology certificate alone reading the clipped reconstruction.

Third, the auxiliary studies are pilots, the scope is 2D, no 3D result
claimed, and the
incremental evidence is exact agreement on the recorded streams: a
local-influence guarantee
and an exactness theorem remain open.

\subsection{Availability}
The released data package accompanies this
submission through the journal's supplementary-artefact channel; the
implementation, versioned source, configurations and provenance will be
deposited in an archival repository with the final publication. All nineteen
benchmarks are public, cited in Section~\ref{sec:Experiment}, and used
unmodified for evaluation only under their licenses. That implementation fixes
the remaining tie resolutions and serves as the executable reference
(Supplementary Appendix~\ref{app:method_derivations}). Together, the
certificates, the fixed operating point and the exact incremental agreement
make skeletonization a dependable, editable primitive.
 
\FloatBarrier

\section*{Acknowledgment}
This work was supported in part by the Defense Advanced Research Projects
Agency (DARPA) under award HR001123S0011.
This work used Bridges-2 at the Pittsburgh Supercomputing Center through
allocation cis220039p from the Advanced Cyberinfrastructure Coordination
Ecosystem: Services \& Support (ACCESS)
program, which is supported by
U.S. National Science Foundation grants \#2138259, \#2138286, \#2138307,
\#2137603, and \#2138296.
The first author thanks Yanhu Zhou, Zongli Tang, Yuming Tang, Zhongrong Ye,
Zhaofang Deng, Yancan Zhou, Weiwei Zhang, Shoujin Zhou, and his family for their
support, and Yunqi Dong, Yongbin Yu, Qing Guan, Yong Liao, Mingxia Sun, Maozhu Wang,
Furi Xiang, Chi-En Teh, and Yunchao Yao for their encouragement.

\ifCLASSOPTIONcaptionsoff
  \newpage
\fi

\bibliographystyle{IEEEtran}

\clearpage

\end{document}


\title{Supplementary Material:\\ SpiderSkel: Sub-Pixel Multi-Tangent 2D Skeletonization from Inscribed-Disc Geometry}
\hypersetup{pdftitle={Supplementary Material: SpiderSkel: Sub-Pixel Multi-Tangent 2D Skeletonization from Inscribed-Disc Geometry},
  pdfauthor={Zhexian Zhou, Lian Ren, Xiao Feng, Sebastian Scherer}}
\author{Zhexian~Zhou,
        Lian~Ren,
        Xiao~Feng,
        and~Sebastian~Scherer,~\IEEEmembership{Senior~Member,~IEEE}}
\markboth{Supplementary Material}{Supplementary Material}
\maketitle
\bstctlcite{IEEEtran:BSTcontrol}

\appendices
\renewcommand{\thesubsection}{\thesection.\arabic{subsection}}
\renewcommand{\thesubsectiondis}{\thesubsection}
\renewcommand{\thefigure}{S\arabic{figure}}
\renewcommand{\thetable}{S\arabic{table}}
\setcounter{figure}{0}
\setcounter{table}{0}
\raggedbottom

\makeatletter
\setlength{\@fptop}{0pt}
\setlength{\@fpsep}{8pt plus 1fil}
\setlength{\@fpbot}{0pt plus 1fil}
\setlength{\@dblfptop}{0pt}
\setlength{\@dblfpsep}{8pt plus 1fil}
\setlength{\@dblfpbot}{0pt plus 1fil}
\makeatother

\section{Evaluation Protocol and Baseline Configuration}
\label{app:evaluation_protocol}

Table~\ref{tab:baseline_configuration} records, for every baseline whose
behaviour its citation does not already determine, the executable adapter, its
software version or repository snapshot, and its run configuration. Every
baseline runs at those documented reference parameters; Appendix~\ref{app:hjs_gamma}
sweeps the one consequential choice, the HJS flux threshold $\gamma$.

\noindent\textbf{Common evaluation wrapper.} Every input is converted to a nonzero foreground mask, black-only outer rows and columns are cropped, and a one-pixel zero border is added. No dataset-specific baseline tuning, resizing, spur pruning, or connectivity repair is applied. An adapter exception is logged with the method and image path, replaced by an all-zero skeleton, and therefore scored as a failure; missing optional dependencies are excluded before evaluation and remain visibly absent from the cache. All outputs are cast to foreground/nonforeground for the shared evaluator. Five baselines are omitted from the table because their citation determines them fully: \texttt{hilditch}, \texttt{k3m}, \texttt{luwang}, \texttt{stentiford} and \texttt{zhangsuen} are our own implementations of the cited algorithms, run at their documented defaults.

\begin{table*}[!t]
\centering
\caption{Baseline implementation audit. Package versions are those of the paper environment; vendored adapters are pinned by repository snapshot \texttt{174362d}. Unless a row says otherwise, the adapter accepts and returns a two-dimensional Boolean mask}
\label{tab:baseline_configuration}
\begin{threeparttable}
\scriptsize
\setlength{\tabcolsep}{2.5pt}
\renewcommand{\arraystretch}{1.10}
\begin{tabular}{p{0.13\textwidth}p{0.33\textwidth}p{0.46\textwidth}}
\toprule
Paper name & Implementation / version & Configuration \\
\midrule
\texttt{lee} & scikit-image 0.24.0, \texttt{skeletonize(method='lee')} & Defaults. \\
\texttt{zhang}\tnote{b} & scikit-image 0.24.0, \texttt{skeletonize(method='zhang')} & Defaults. \\
\texttt{medial\_axis} & This work's deterministic C++ port of scikit-image 0.24.0 \texttt{medial\_axis} & Reference defaults, with a fixed raster-index tie-break in place of the reference's PRNG shuffle over equal-key pixels (\texttt{rng=None}), so the scored skeleton is reproducible. \\
\texttt{thinning} & scikit-image 0.24.0, \texttt{thin} (Guo--Hall) & Defaults; same algorithm as \texttt{guo-hall}, independent implementation. \\
\texttt{sitk} & SimpleITK 2.5.3, \texttt{BinaryThinning} & Defaults; array round trip. \\
\texttt{guo-hall} & OpenCV 4.11.0.86, \texttt{ximgproc.thinning} & Input scaled to $\{0,255\}$; \texttt{THINNING\_GUOHALL}. \\
\texttt{mahotas} & Mahotas 1.4.18, \texttt{thin} & Defaults. \\
\texttt{kimimaro} & Clean-room C++ TEASAR at the kimimaro 5.8.1 2D defaults & \texttt{anisotropy=(1,1,1)}, \texttt{dust\_threshold=0}, no progress; graph edges rasterized. TEASAR is natively 3D, so this row is the declared 2D adapter. \\
\texttt{dsepruning} & This work's deterministic C++ port of dsepruning 0.1.0 \texttt{skel\_pruning\_DSE} & \texttt{min\_area\_px=100}, run over the deterministic \texttt{medial\_axis} above rather than the reference's randomised one, so no tie-order nondeterminism remains. \\
\texttt{aof} & Vendored average-outward-flux adapter & 60 samples, $\epsilon=1$, flux threshold 18, final thinning enabled. \\
\texttt{hjs} & Vendored HJS adapter & $\gamma=2.5$. \\
\texttt{afmm+tft}\tnote{a} & AFMM reference implementation, method~2, through its own \texttt{pyafmm} 0.1.0 binding + scikit-image 0.24.0 & \texttt{skeletonize(mask,35.0,False)} then scikit-image thinning. \\
\texttt{afmm}\tnote{a} & AFMM reference implementation, method~1 & Threshold 5.0; reproduces the reference implementation bit for bit. \\
\texttt{voronoi} & Vendored SciPy Voronoi adapter & Boundary spacing 3 px; chord residual threshold 3; minimum branch 10; no post-thin. \\
\mbox{\texttt{MATLAB Voronoi}} & Vendored MATLAB-aligned Voronoi adapter & Trim 1.5; full boundary sampling (\texttt{fast=-1}); rasterized edge output. \\
\texttt{unet} & Released vendored checkpoint; separately retrained protocol & Released probability map thresholded by its adapter; retrained results use Appendix~\ref{app:learned_comparison}. Off-the-shelf format mismatch is reported separately from held-out retraining. Input resized to $256\times256$ (nearest-neighbour) for inference and the prediction resized back to native size. \\
\texttt{skelite} & Released \texttt{model\_000400.pt}; separately retrained protocol & Released DRIVE configuration, scale 2 and binary threshold 0.5; retrained results use Appendix~\ref{app:learned_comparison}. Empty or near-empty released outputs are retained as failures, not silently repaired. \\
\bottomrule
\end{tabular}
\begin{tablenotes}[flushleft]\scriptsize
\item[a] Both AFMM rows come from the same upstream, the AFMM reference implementation: \url{https://github.com/preethamam/AFMM-Image-Morphological-Skeletonization} (MIT).
\item[b] \texttt{zhangsuen}, scored as a separate method, is our own independent implementation of the same algorithm.
\end{tablenotes}
\end{threeparttable}
\end{table*}
 
The primary protocol is \emph{whole-image}: a method runs once on the entire
canvas, with object-aware decoding and structural evaluation and the
reconstruction terms pooled over the whole image. The componentwise alternative
runs the method on one object at a time, in its own bounding-box crop with the
neighbouring objects erased; it is the protocol used for the substitution
ablation of Section~\ref{subsubsec:component_substitution}, the robustness
study, the threshold sweep, the learned-model comparison, the failure mining
and score-composition studies, and the deployment decomposition of the
incremental engines. Implementations disagree about foreground that meets the
image frame, reading it as a shape bound (the thinning family) or as a
field-of-view bound (a distance-based medial axis); the evaluation removes that
ambiguity identically for every method by cropping black-only outer rows and
columns and adding a one-pixel background border to every input. The protocol
choice does not favour the methods proposed here: under the
translation-equivariant probe frame the three SpiderSkel variants produce
identical skeletons under both protocols on all $11{,}323$ images, a corpus
property rather than a general equivalence since erasing a neighbour inside a
crop margin changes the distance transform, and the largest single change is
the weakest baseline improving by $1.86$~pp under the primary protocol.

\subsection{Robustness protocol}
\label{app:robustness_full_tables}

The consolidated robustness table of Section~\ref{subsec:robustness}
(Table~\ref{tab:robustness_skelqs}) and its per-category view
(Fig.~\ref{fig:robustness_per_category}) use a subsample fixed before
analysis that spans all nineteen benchmarks: every $k$-th image per dataset
with $k=\max(1,\lfloor n/200\rfloor)$, so the largest datasets are capped
near $200$ images and every small dataset contributes in full, $2{,}715$
images in total. The roster is the consolidated table's, the three SpiderSkel
variants and the six strongest baselines by Overall score; the proposed
methods hold $(\phi_l,\phi_h)=(0,13\pi/16)$ fixed across all datasets and
severities, no degraded robustness instance is used to select the operating
point, and every
cell, degraded and clean alike, is scored under the componentwise protocol of
Appendix~\ref{app:evaluation_protocol}. \emph{Boundary perturbation} adds a
band-limited Gaussian random field to the signed-distance function (amplitude
$0.5/1.0/2.0$, smoothing $\sigma=3/5/7$ pixels); \emph{dilation} applies
$k\in\{1,3,5\}$ iterations of a fixed $3\times3$ kernel, an effective
$(2k{+}1)^2$ structuring element; \emph{resolution degradation} is $2/4/8\times$
area downsampling followed by nearest-neighbour upsampling; \emph{rotation} is
nearest-neighbour resampling through $15^\circ/30^\circ/45^\circ$ about the
canvas centre with the output canvas expanded to the rotated bounding box, so
that no foreground is ever clipped and the residual degradation is exactly
the lattice resampling. The per-category cells for all twelve conditions and
every method in the roster accompany the released package as
\texttt{robustness\_per\_category.csv}, on the same dataset-macro basis as the
figure.
\section{Failure Analysis, Score Soundness, and Threshold Selection}
\label{app:failure_analysis}
\label{app:failure_cases}%

Fair evaluation is not equal behaviour. Decomposing both certificates per
method over all $11{,}323$ images and $31{,}209$ objects separates what a
rounded rate cannot: hjs and stentiford both print $T_{\mathrm{recon}}=0.99$
(Table~\ref{tab:category_avg_skelqs_em}), yet one fails on one image, the
other on $82$. Across the nine methods whose skeleton certificate is clean but
whose reconstruction certificate is not, the spurious reconstruction hole is
the only mode that occurs, reflecting how those thinners
fail; among the thirteen failing both certificates, \texttt{medial\_axis}
fails almost entirely by introducing cycles the object lacks ($3{,}046$
images, $26.9\%$, through $25{,}628$ extra-hole instances). Off-object skeleton
mass cannot affect the certificate: the decoder discards it before structural
evaluation. Under the whole-image protocol
that discard is real: the baselines place measurable skeleton mass off-object,
our three methods none. The omission term captures a different failure,
reconstruction locally accurate wherever it exists yet omitting whole interior
structures; Appendix~\ref{app:hjs_gamma} records its threshold sensitivity.

We mine our failures exhaustively and prospectively, pairing all $11{,}323$
images between the SpiderSkel variants and four comparison baselines
(\texttt{medial\_axis}, \texttt{zhang}, \texttt{guo-hall}, \texttt{lee}); a
\emph{strict} failure is a variant losing on the constituent \emph{and}
composite SkelReconQS against the same baseline, with fixed tolerance bands
(strict sub-pixel reconstruction Dice more than $0.5$~pp behind the best
comparison baseline; Reduction Rate more than $2$~pp behind the best thinning
baseline). No strict topology or reconstruction failure occurs. The one
surviving trade-off is skeleton density: strict Reduction failures on $8$
images for SpiderSkel-N8 ($0.07\%$, worst gap $4.0$~pp), $8$ for
SpiderSkel-DualC and $64$ for SpiderSkel-Scaffold ($0.57\%$, worst gap
$16.8$~pp), concentrated on silhouette shapes near the resolution limit plus
isolated crack masks. They occur where the multi-tangent neighbourhood
preserves several near-equidistant medial candidates along the same ridge; the
retained pixels remain centres of exactly inscribed discs, so the certificates
and the reconstruction terms are unaffected, and the optional $+R_2$ reduction
trims part of the residual density.

\subsection{Soundness of the composite score}
\label{app:score_soundness}
\label{app:score_composition}

On the nonzero support (missing-object gate passed), $\log\mathrm{SkelReconQS}$
is exactly the sum of six additive log-terms, each factor's logarithm scaled
by its exponent; contributions and rank
sensitivities are measured there, further restricted to images with at least
eight of the $25$ methods scoring above zero, with vetoed and zero cells
counted, not imputed. Compaction and the two
reconstruction-fidelity terms $F$ and $G_{\mathrm{avg}}$ are near-saturated
across competitive methods: each carries under $1\%$ of the top-10
between-method variance, and dropping any leaves the 25-method ranking
essentially unchanged ($\rho\geq0.996$). Which term discriminates is
category-dependent: connectivity
carries $88\%$ of between-method variance on solid shapes, the omission factor
$87$--$89\%$ on the three thin-structure categories. Holding omission at $1$
drops the correlation to $\rho=0.907$ with hjs first; every other single-term
drop preserves the SpiderSkel top-three set. Omission alone is gameable by
under-compaction: a dense predictor keeping $40\%$ of the object's pixels
still attains a perfect omission factor, and removing it compresses the top
fifteen methods into a $2.8$-point band. Across all $729$ positive exponent
reweightings the three SpiderSkel variants remain the top-three set,
structurally exact on $11{,}323/11{,}323$ images; the order within the set is
not separated, SpiderSkel-N8 and SpiderSkel-DualC differing by $0.004$~pp in
Overall, inside the confidence half-width.

\subsection{Threshold selection and hyperparameter sensitivity}
\label{app:threshold_search}

Section~\ref{sec:Experiment} fixes the operating point
$(\phi_l,\phi_h)=(0,13\pi/16)$, the SkelReconQS argmax of the $81$-cell sweep
over the evaluation corpus (all $11{,}323$ images, SpiderSkel-N8 chain,
dataset-macro within category then averaged over the four categories); the
argmax is one of only six cells whose topology certificate is exact on every
image. A leave-one-category-out check scores each held-out
benchmark category at the argmax of the other three under the same equal-weight
macro. Three of the four folds select the reported pair exactly, losing at most
$0.01$~pp against that category's own best cell; the fourth, holding out
document binarisation, selects $(0,15\pi/16)$, costing $1.60$~pp. Selecting on
the corpus therefore buys at most about $1.6$~pp on one of four categories and
$0.40$~pp on average, an upper bound on the selection effect rather than a
correction applied to any reported number.

The six non-$\phi$ hyperparameters (the support-sample count
$N_{\mathrm{s}}$, the probe decay angle $\phi_d$, the representative-retention
cap $N_{\max}$, the span-comparison rounding precision, the number of angular
thinning bins and the optional closing reduction's cap
$\tau_{\max}$) were swept one at a time around their defaults on $135$ images
from CrackForest, DIBCO, DRIVE, DeepGlobe and MPEG7, $\phi$ at the operating
point, every skeleton scored under the whole-image protocol. Every default
reproduces the reported skeleton exactly and sits on a plateau: no swept
alternative gains more than $0.03$~pp, and $T_{\mathrm{skel}}$,
$T_{\mathrm{recon}}$ and $\mathrm{Conn}$ are exactly $1.000$ at every swept
value. A stress sweep maps the boundaries: thin-structure reconstruction
collapses once $N_{\mathrm{s}}$ falls below about three, with
$N_{\mathrm{s}}{=}1$ the one setting in either sweep whose
certificate leaves exactness ($T_{\mathrm{recon}}=0.9997$); $\tau_{\max}$ above
its default over-prunes the optional reduction by more than $75$~pp on DRIVE;
the cap $N_{\max}\in\{2,5,10,20\}$ moves the macro score by at
most $0.01$~pp while changing individual outputs on $69$ of $135$ images,
per-image range at most $0.13$~pp; and $\phi_d$ is flat from $\pi/4$ to
$3\pi/4$ within $0.07$~pp. These certificate results are empirical properties
of the sweeps, not guarantees for unseen masks or parameter values.
\section{Verified Incremental Engines for the SpiderSkel Variants}
\label{app:incremental_algorithm}

Each engine targets one of the three no-reduce SpiderSkel variants end-to-end,
outputting the \emph{entire} method's skeleton, not a sub-pipeline; its
locality is an observed performance property, not an assumption, since
distance-field influence can be global. Per
Appendix~\ref{app:evaluation_protocol}'s componentwise protocol, each
foreground object owns its crop, core state and skeleton, and two separately
accounted mechanisms produce the speedup. \emph{Retention} normalises an edit
to the pixels whose Boolean values toggle, and objects whose extent the toggles
do not intersect are retained, their skeletons reused unchanged. \emph{The
persisted field} keeps the core's candidate field resident for a touched object
and re-derives its skeleton through the same shared core, replaying only what
the edit invalidates, with a per-object complete recomputation when the
persisted state cannot certify the edit. As paired medians over the fixed
five-stratum $240$-image sample, retention alone gives $27.06\times$,
$33.62\times$ and $21.38\times$ on canvas edits for SpiderSkel-N8,
SpiderSkel-DualC and SpiderSkel-Scaffold, the field adding no measurable
factor; on foreground edits retention gives $1.00\times$ and the field path,
carrying $99.3\%$ of the touched objects, $1.38\times$, $2.00\times$ and
$1.50\times$. Edit streams are fixed lists from a seeded generator, each edit
toggling exactly one pixel: \emph{foreground} edits uniform on the original
mask's foreground, \emph{canvas} edits uniform on the padded canvas so most
steps touch no object; $25$ steps per workload per image, the two workloads
never pooled. Within a step the incremental update and complete recomputation
run back-to-back in seeded random order, cancelling machine drift inside the
pair; the bootstrap resamples images, not steps. Each recorded step's toggles
are applied to both the persistent engine and a fresh complete run of the
method, and exact equality of the two skeletons is recorded: $23{,}040$ steps,
zero mismatches (the verification pass walks its own $16$-step streams). Per edit the work
stays linear in the touched object's crop plus replay proportional to the
invalidated region, so an edit changing an object's partition approaches one
complete recomputation of it; Table~\ref{tab:incsafe_paired}'s speedups
describe the evaluated hardware and trace distribution, not worst-case bounds,
and any failure of the persisted structure is recoverable by re-initialising from the
current image.
\section{Overall Comparison: Trade-off Reading, Runtime, and Baseline Retuning}
\label{app:overall_full}

The Reduct.\ and Dice$_{\mathrm{sp}}$ columns of main-text
Table~\ref{tab:category_avg_skelqs_em} report per-category Reduction Rate
and strict sub-pixel reconstruction Dice for the same twenty-five methods as
its SkelReconQS cells. The Pareto claim, stated in the (Dice$_{\mathrm{sp}}$,
Reduction Rate) plane of main-text Eq.~\eqref{eq:dice_sp}, is evaluated only
inside each category, because the two metrics' baselines vary substantially with
foreground geometry: thin retinal vessels yield much lower Reduction Rates
than solid shape blobs, and the achievable Dice ceiling on heavily-textured
masks differs from that on smooth shapes. The per-method decomposition of the
three structural quantities, with the full $25\times19$ per-dataset grid,
accompanies the released package as \texttt{overall\_per\_dataset.csv}.

\subsection{Runtime details}
\label{app:runtime}

Section~\ref{subsec:runtime}'s runtime protocol ran one uniform
single-threaded C++ benchmark (gcc 13.2.1, no fast-math,
\texttt{OMP\_NUM\_THREADS}${}={}1$) on exclusive PSC Bridges-2 RM nodes over
the stratified $16$-image set, four per category. Each occasion took $11$
repeats per image, the first discarded as warm-up; reported cells are
medians of three independent occasions, macro-medianed within then
across categories. Under this protocol SpiderSkel-Scaffold, SpiderSkel-DualC and
SpiderSkel-N8 take $0.062$, $0.078$ and $0.083$~s end-to-end at
$\mathcal{O}(N{+}T)$, against $0.012$~s for the fastest thinner, \texttt{zhang};
per-method, per-category medians for both scopes (end-to-end, core
skeletonization-only) and a spread column (median per-image interquartile
range of retained repeats) accompany the released package as
\texttt{runtime\_per\_category.csv}.

\subsection{Baseline threshold sensitivity}
\label{app:hjs_gamma}

Appendix~\ref{app:evaluation_protocol} keeps every baseline at its documented
reference parameters. Retuning the HJS flux threshold $\gamma$ narrows the gap
without reversing it: increasing $\gamma$ retains more branches, which
reduces omission at the cost of a lower Reduction. At $\gamma{=}5.0$, twice its
reference default, HJS narrows the sample-level SkelReconQS gap to SpiderSkel-N8 from
$9.0$~pp to $0.15$~pp on a 465-image stratified sample,
but on that same sample its Reduction falls from $0.905$ to $0.896$, raw
reconstruction Dice is essentially unchanged ($0.931$ to $0.930$; the sweep
records raw Dice, not $\mathrm{Dice}_{\mathrm{sp}}$), and $14$ of the $465$
images regress by more than $1$~pp of SkelReconQS, the worst by $1.7$~pp, so
$\gamma{=}5.0$ is not a uniformly superior operating point.

On the full corpus, even after this retuning, HJS remains slightly below
SpiderSkel-N8 on both quantities, $0.909$ against $0.911$ SkelReconQS and
$0.901$ against $0.906$ Reduction, and it retains severe omission cases: $27$
images still score $G_{\mathrm{om}}{<}0.9$, the lowest at $0.27$, against
none for SpiderSkel-N8 under the same protocol. Both methods are scored in the same sweep, whose
Reduction pools pixels within each dataset before the macro, so these
magnitudes are comparable with each other but not with the per-image
Reduction Rate of Table~\ref{tab:category_avg_skelqs_em}.

Like the SpiderSkel operating point (Appendix~\ref{app:threshold_search}),
$\gamma{=}5.0$ is selected by sweeping on the evaluation corpus; it is
reported as sensitivity analysis, and the main-table row keeps the documented
default. The sweep over all seven
thresholds $\gamma \in [1.5, 5.0]$, as per-dataset and overall aggregates,
accompanies the released package as \texttt{hjs\_gamma\_sweep.csv}.
\section{Comparison with Trained Learned Skeletonizers}
\label{app:learned_comparison}

\begin{table*}[!t]
\centering
\footnotesize
\setlength{\tabcolsep}{4pt}
\renewcommand{\arraystretch}{1.1}
\caption{Learned-skeletonizer comparison under SkelReconQS on the held-out shape test split ($N{=}1208$ images, twelve datasets; one frozen split and preprocessing for every method). SkelReconQS and its terms use the common integer decoder for every method; only Dice$_{\mathrm{sp}}$ decodes SpiderSkel at its sub-pixel centres. Best per column in bold}
\label{tab:sp3_srqs}
\begin{tabular}{@{}lcccccc@{}}
\toprule
Method & SkelReconQS & $T_{\mathrm{skel}}$ & $T_{\mathrm{recon}}$ & $\mathrm{Conn}$ & Reduct. & Dice$_{\mathrm{sp}}$ \\
\midrule
\textbf{SpiderSkel-N8} & \textbf{96.59} & \textbf{1.000} & \textbf{1.000} & \textbf{1.000} & 94.99 & 99.77 \\
\textbf{SpiderSkel-DualC} & 96.59 & \textbf{1.000} & \textbf{1.000} & \textbf{1.000} & 94.99 & 99.77 \\
\textbf{SpiderSkel-Scaffold} & 96.50 & \textbf{1.000} & \textbf{1.000} & \textbf{1.000} & 94.70 & \textbf{99.80} \\
\midrule
Medial axis & 93.88 & 0.206 & \textbf{1.000} & \textbf{1.000} & 92.58 & 99.59 \\
Zhang & 90.14 & \textbf{1.000} & 0.999 & \textbf{1.000} & 96.09 & 98.02 \\
Lee & 86.88 & \textbf{1.000} & \textbf{1.000} & \textbf{1.000} & \textbf{96.47} & 97.15 \\
Thinning & 88.69 & \textbf{1.000} & 0.999 & \textbf{1.000} & 96.26 & 97.51 \\
\midrule
Skelite (retrained on reference annotations) & 20.05 & 0.062 & 0.927 & 0.228 & 90.95 & 98.91 \\
Attention U-Net (trained on reference annotations) & 29.83 & 0.117 & 0.875 & 0.361 & 91.69 & 97.88 \\
\bottomrule
\end{tabular}
\end{table*}

Table~\ref{tab:sp3_srqs} compares the training-free method with two learned
skeletonizers on the twelve shape-matching benchmark datasets of
\cite{DBLP:journals/ijcv/YangISGKBYG24}: Skelite and an attention U-Net, each
retrained in domain from scratch, not evaluated as released, under its own
authors' loss and optimiser,
best-validation checkpoint retained. The $6{,}030$ paired shape images are
split by image, $80\%$ training / $20\%$ test, stratified so every dataset
appears in the test split, entirely held
out of training and shared by both learned models. Every method is evaluated
at a common $256\times256$ resolution, the learned models' training regime,
so absolute scores are not comparable to
Table~\ref{tab:category_avg_skelqs_em}'s native-resolution values.
SpiderSkel's three variants run componentwise (native mode, no-reduce
definition); SkelReconQS and its terms use the common integer decoder and the
object's own distance-transform radii for every method, only
Dice$_{\mathrm{sp}}$ decoding SpiderSkel at its reported sub-pixel centres;
learned-model binarisation thresholds are chosen on a validation split carved
as $10\%$ of the training portion, leaving the held-out test split untouched;
no thinning applied. Both learned models
collapse through \emph{connectivity}, not fidelity: strict sub-pixel Dice
near $98$--$99$ while $\mathrm{Conn}$ falls to $0.23$ (Skelite) and $0.36$
(attention U-Net); only
SpiderSkel's three variants and \texttt{lee} are exact on all three
structural quantities on every held-out image.
\section{Method Derivations}
\label{app:method_derivations}

This appendix complements the Method (Section~\ref{sec:Proposed Method})
with the tie-aware envelope (\ref{app:envelope}), two tangent-point
geometry facts (\ref{app:tangent_basics_full}), the incircle modeling's
exactness, ranking and bounds (\ref{app:incircle_full}), and the closing
stages of SpiderSkel-N8 (\ref{app:closing_stages}). Wherever a rule
leaves a choice open (scan order, equal keys, equal-length paths), the
implementation resolves it in a fixed deterministic order and serves as
the executable reference for such details.

\subsection{Tie-aware exact-distance envelope}
\label{app:envelope}
\label{app:envelope_extension}

Building on the tie-aware separator of
Section~\ref{subsec: Feature-aware Envelope Calculation}, this subsection
states how the query pass reads and bounds the ties it exposes, and at
what cost. Two sets are involved: the \emph{full tie pool} $\mathcal{Q}_P$
of Eq.~\eqref{eq:tangent_set}, on which the grid-point span $S_P$ is
always read, and the \emph{capped representative set} that Stage~3 uses
for sub-pixel interpolation only.

\paragraph{Full tie pool and exact span}
At a foreground pixel $P$ the query pass admits an adjacent envelope range
exactly when its evaluated squared distance equals the minimum, so it
collects all of $\mathcal{Q}_P$. A single streaming pass over the pool tracks the largest
circular gap and its two flanking sites in constant state, so the span
$S_P$ of Eq.~\eqref{eq:span} is exact for every pool size and the pair
$(A^{\ast},B^{\ast})$ realising it is exported, without the pool ever
being sorted or stored.

\paragraph{Capped representative set: interpolation only}
Wherever Stage~3 would read $\mathcal{Q}_P$ it reads instead a
representative subset $\mathcal{Q}^{\mathrm{rep}}_P$ of at most $N_{\max}$
sites ($20$ in every reported run). The adopted rule,
\emph{bounded-streaming retention}, keeps every site when the pool fits
the cap; otherwise it divides the pool into $N_{\max}$ equal angular
buckets about $P$, keeps the first-arriving site of each non-empty
bucket, and fills the remaining slots with the earliest unkept sites
($S_P$ is independent of the bucket origin). It costs at most two further
linear sweeps of the pool, one when the pool fits the cap, and $N_{\max}$
words of state.
The alternative, \emph{sorted-pool retention}, materialises the pool and,
above the cap, sorts and trims it by the greedy angular thinning of
Appendix~\ref{app:incircle_full}. The two rules coincide whenever the
pool fits the cap, which holds at every pixel of the evaluation corpus,
so they produced identical skeletons on every evaluation image; above the
cap they can select different representatives, and there the streaming
rule stays linear while the trim is quadratic in the worst case.
The runtime measurements (Appendix~\ref{app:runtime}) were taken under
bounded-streaming retention.

\paragraph{Complexity}
For $N$ pixels ($N$ as in Table~\ref{tab:prior_art_claims}), $N_f$ of
them foreground, the two separable envelope scans are $O(N)$; the dense
fields use $O(N)$ memory and the foreground-compact candidate array
$O(N_fN_{\max})$. Span and retention together cost $O(N+T)$, where
$T=\sum_P(|\mathcal{Q}_P|-1)$ over the foreground pixels is the total tie
mass beyond one site per pixel. Each $|\mathcal{Q}_P|$ is bounded by the
number of lattice points on the circle of squared radius $d(P)^2$ (the
sum-of-two-squares function), so $T$ is not a fixed constant in the worst
case, though it measures $T\approx0.17N_f$ on a census of $523$
evaluation-corpus images.
\subsection{Tangent-point geometry: two facts}
\label{app:tangent_basics_full}

Two geometric facts underlie the joint evaluation of
Section~\ref{sec:Tangent Points Geometric Basics}: along an inward normal
the contact transition is a single cut position, and digital extension
from a boundary site is set-valued.

\paragraph{Fact 1: two-phase normal-line distance, a single cut position}
Idealise $\partial\Omega$ as a polygonal curve through the ring pixels,
$d$ its distance, with inward normal $\boldsymbol{\nu}$ at
$Q\in\partial\Omega$ (any direction of the inward normal cone at a corner)
and the ray $q(t)=Q+t\boldsymbol{\nu}$, $t\ge0$. The open balls of radius
$t$ centred at $q(t)$ are nested and increasing in $t$, all touching the
boundary at $Q$, so the set of $t$ at which $Q$ is still a nearest
boundary point of $q(t)$ is an interval
$[0,t_{\mathrm{cut}}(Q;\boldsymbol{\nu})]$:
\begin{equation}
d(q(t))=t\ \ (t\le t_{\mathrm{cut}}),\qquad
d(q(t))<t\ \ (t>t_{\mathrm{cut}}),
\end{equation}
a different boundary point being closer beyond the cut. On the pixel grid
both $Q$ and $\boldsymbol{\nu}$ are quantised, so the cut is perturbed and
a single interior sample along one ray cannot locate it; candidates are
therefore evaluated jointly over a neighbourhood.

\paragraph{Fact 2: digital extension is set-valued}
The grid-aligned step
$\boldsymbol{u}=\boldsymbol{\nu}/\|\boldsymbol{\nu}\|_\infty$ induces the
digital ray $Q_k(\boldsymbol{u})=Q+\operatorname{round}(k\boldsymbol{u})$,
$k\in\mathbb{Z}_{\ge0}$, rounded componentwise. Any such ray reaching a
grid point $P$ does so at $k=\|P-Q\|_\infty=\delta_{\max}$, its main-axis
entry being $\pm k$; what is not unique is the direction and the sample
after $P$. On an axis-aligned edge the normal, hence the ray, is unique,
whereas at a concave corner (including the concave staircase corners of a
digitised diagonal) the inward normals span an arc, $\pi/2$ at a right
angle, so distinct steps reach the same $P$ and part at
$Q_{k+1}(\boldsymbol{u})$, which varies on the secondary and
zero-difference axes. Appendix~\ref{app:incircle_full} bounds that
increment per axis. The stencil $\mathcal{P}^P_Q$ of
Eq.~\eqref{eq:ga_neighbourhood} is a finite grid-aligned selection of
these continuations; a single-direction estimate would discard the
geometric information the fan carries.
\subsection{Incircle tangent-point modeling: exactness, ranking and bounds}
\label{app:incircle_full}

This subsection gives the deferred algebra of the 2D method; the 3D
counterparts are not reproduced, since no claim depends on them.

\paragraph{Proof of the closed-form interpolation (main-text
Lemma~\ref{lem:two_site_interp})}
With $\|NX\|^2=\|(P-X)+e\|^2$ for $X\in\{A,B\}$,
\begin{equation}
\begin{aligned}
\|X\tilde{P}(p)\|^2&=\|(P-X)+pe\|^2\\
&=(1-p)\|PX\|^2+p\|NX\|^2+p(p-1)\|e\|^2;
\end{aligned}
\end{equation}
equating the two cancels $p(p-1)\|e\|^2$ and leaves a linear equation in
$p$ whose solution is Eq.~\eqref{eq:p_closed}. \hfill$\square$

\paragraph{Nearest-shell exactness}
Each admissible bucket $G_p$ of the main text meets at $\tilde P=P+p\,e$;
its skip condition $\|PB\|\neq d(P)$ gives $\Delta_P>0$, so the lemma's
hypothesis holds for every pair. With $p=p_n/p_d$ for the integers
$p_n=\Delta_P$ and $p_d=\Delta_P+\Delta_N>0$ of Eq.~\eqref{eq:p_closed},
$p_d^2\|\tilde P-X\|^2$ is an integer polynomial in the grid data for
every participating site $X$, so the nearest shell over the bucket's
participating set $\mathcal{Q}^{(p)}$ and membership of it are found by
exact integer comparison, the common positive factor $p_d$ divided out.
A single-pair group is co-radial by the lemma and needs no
retention step. Two further caveats: the retention cap of Appendix~\ref{app:envelope} can withhold
minimisers, and the two sites realising the span on the retained shell may
both come from $\mathcal{Q}_P$ or both from $\mathcal{Q}_N$.

\paragraph{Bucket identity and gcd safety}
Expanding $\Delta_P+\Delta_N$ with $\|NX\|^2=\|(P-X)+e\|^2$ cancels every
squared norm:
\begin{equation}
p_d=\Delta_P+\Delta_N=2\,e^{\top}(B-A),
\end{equation}
an integer, so the bucket key $(p_n,p_d)$ reduced by $g=\gcd(p_n,p_d)$
names the sub-pixel point exactly. Reduction scales the $p_d$-divided
pairwise differences of the shell comparator, homogeneous of degree one in
$(p_n,p_d)$, by $1/g$ and leaves their signs, the only thing consulted,
untouched, so it changes no decision.

\paragraph{Ranking}
Surviving candidates are ranked by decreasing support evidence, then
decreasing span rounded to one decimal, then increasing realised
displacement $\|p\,e\|$, remaining ties broken by a fixed order on the
tangent-site coordinates; the winner is then subject to the main text's
replacement rule.

\paragraph{Integer-translation equivariance}
Every quantity above depends on grid positions only through the
differences $e=N-P$, $X-P$ and $X-N$, or through coordinate comparisons
that a common translation preserves, so an integer translation $t$ sends
$\tilde P\mapsto\tilde P+t$ and leaves $p$ and the span descriptors
unchanged; evaluated in the frame of $P$ the operands are small integers
and invariant scalars, so this also holds in floating point, and a
translation clear of the frame translates exactly.

\paragraph{Sorted-pool retention: greedy angular trim}
When the pool $\mathcal{Q}_P$ exceeds the cap under sorted-pool retention
(Appendix~\ref{app:envelope}), its sites are sorted by polar angle about
$P$ and the site whose two neighbouring cyclic gaps sum to the smallest
cross-angle is deleted and its gaps merged, repeatedly, until $N_{\max}$
sites remain; the survivors form $\mathcal{Q}^{\mathrm{rep}}_P$, while
$S_P$ is still read on the full pool. Bounded-streaming retention never
invokes the trim. Including the sort, the trim costs
$O(n\log n+n(n-N_{\max}))=O(n^2)$ in the worst case for a pool of $n$
sites, with $O(n)$ temporary storage, against $O(n)$ for the streaming
rule; downstream storage and work stay bounded by $N_{\max}$.

\paragraph{Per-axis increment of the next sample}
Let $\delta=P-Q\neq0$, $d_i=|\delta_i|$ and $d^*=\delta_{\max}$, and let
a continuous direction $\boldsymbol{u}$, with
$\operatorname{sgn}u_i=\sigma_i$ where $\delta_i\neq0$, realise $P$ at
step $d^*$, i.e.\ $\operatorname{round}(d^*|u_i|)=d_i$, equivalently
$d_i-\tfrac12\le d^*|u_i|<d_i+\tfrac12$, on every axis. The increment of
the next sample is $\Delta_i=\operatorname{round}((d^*+1)|u_i|)-d_i$, and
$(d^*+1)|u_i|=d^*|u_i|+|u_i|$. On a secondary axis, $1\le d_i\le d^*-1$
gives $0<|u_i|<1$, hence
\begin{equation}
(d^*+1)|u_i|\in\bigl(d_i-\tfrac12,\ d_i+\tfrac32\bigr)
\quad\Longrightarrow\quad \Delta_i\in\{0,1\}.
\end{equation}
The three axis classes are:
\begin{center}
\footnotesize
\setlength{\tabcolsep}{4pt}
\begin{tabular}{@{}lll@{}}
\toprule
Axis class & Constraint on $u_i$ & $\Delta_i$ \\
\midrule
secondary, $1\le d_i\le d^*-1$ & $0<|u_i|<1$ & $\{0,1\}$ \\
primary, $d_i=d^*$ & $u_i=\sigma_i$ (model) & $\{1\}$ \\
zero-difference, $d_i=0$ & $|u_i|<1/(2d^*)$ & $\{-1,0,1\}$ \\
\bottomrule
\end{tabular}
\end{center}
The primary row holds because the extension model fixes $u_i=\sigma_i$ on
every main axis (a tied diagonal displacement steps all of them at once,
Section~\ref{subsec: Geometry-adaptive Neighbourhood}), pinning
$\Delta_i=1$ where the rounding bound alone would allow $\{0,1,2\}$; the
zero-difference row follows from $|u_i|<\tfrac1{2d^*}$, which gives
$|(d^*+1)u_i|<1$. These per-axis sets are what make $\mathcal{P}^P_Q$
(Eq.~\eqref{eq:ga_neighbourhood}) non-singleton.
\subsection{Closing stages of SpiderSkel-N8}
\label{app:closing_stages}

Section~\ref{subsec: Algorithm Pipeline} summarises Stage~6 and the closing
reduction; here each closing stage is stated as SpiderSkel-N8 executes it,
once on the whole input canvas under the primary protocol
(Appendix~\ref{app:evaluation_protocol}). $\Omega$ is the input foreground
and $S$ the evolving skeleton set; a rule applied by \emph{sweeps} is
iterated over the canvas until nothing changes, each change taking effect
immediately for the pixels visited after it. Steps~1--6 produce the
SpiderSkel-N8 skeleton; step~7 is run only in the optional $+R_2$ configuration.

\begin{enumerate}
\item \emph{Interior filling.} After the relocation of Stage~5, every
  foreground pixel whose four axial neighbours all lie in $S$ is added, to
  a fixpoint. The rule only adds and an addition never disables another,
  so scan order does not matter; no other fill runs.
\item \emph{Cleaning.} Within each 4-connected part of $\Omega$, only the
  champion among the 8-connected fragments of $S$ is kept, the fragment
  attaining the largest EDT value (ties: more pixels, then a fixed order);
  the other fragments are deleted outright, with no topology test, and
  nothing is added.
\item \emph{Bridging.} Within each 8-connected object of $\Omega$, the
  fragments left by cleaning are joined into one component by adding, one
  fragment at a time, the interior of a shortest 8-connected path through
  the object from the already-connected fragments to the nearest
  unconnected one. No cycle test is applied; a pure-foreground cycle a
  bridge closes is removed in step~5.
\item \emph{Hole encirclement.} Steps~4--5 use the $(8,4)$ pairing of
  step~7: an object hole is a 4-connected background component enclosed
  by $\Omega$, and the skeleton encloses it when $S$ 4-separates it from
  the exterior. For each hole the skeleton does not yet enclose (checked
  against $S$ as augmented by earlier holes), the ring of foreground
  pixels 8-adjacent to the hole is added and, if that ring does not touch
  the existing skeleton, joined to it by a shortest path through the
  object. The additions are discarded if they increase the number of
  skeleton components.
\item \emph{Spurious-cycle fill.} Every bounded 4-connected region of the
  skeleton's complement that contains no background pixel is a spurious
  cycle and is filled in full; a region containing background encloses a
  real hole and is kept.
\item \emph{Thinning.} Let $d$ be the EDT of $\Omega$,
  $r(P)=\lceil d(P)\rceil$ the decode radius (not the support radius
  $r_\ast$), $\mathcal{B}(P)$ the closed disc of radius $r(P)$ about $P$,
  $c(q)$ the number of current skeleton discs
  containing $q$, $\widehat S_P$ the selected span of Stage~3 (zero where
  fewer than two tangent labels are held), $\Theta_k=[k\pi/4,(k+1)\pi/4)$,
  $k=0,\dots,7$, its eight bins, $\Lambda$ the distinct values of $d^2$
  over the pre-thinning skeleton in ascending order, $n_4(P)$ the number
  of axial neighbours of $P$ in $S$, $N_8(P)$ its eight ring pixels and
  $C_8$ the number of 8-connected components. For each level
  $\lambda\in\Lambda$ in ascending order and each bin $k$ in turn, a sweep
  deletes $P$ (updating $c$) iff
  \begin{align}
  &d^2(P)\le\lambda,\quad \widehat S_P\in\Theta_k,\quad n_4(P)\in\{2,3\},\nonumber\\
  &C_8\bigl((N_8(P)\cup\{P\})\cap S\bigr)=C_8\bigl(N_8(P)\cap S\bigr),
  \label{eq:thin_rule}\\
  &\nexists\,q\in\mathcal{B}(P)\cap\Omega:\ c(q)=1,\nonumber
  \end{align}
  that is, the pixel is neither an end nor an interior pixel, its removal
  keeps the local component count, and it is not the sole cover of any
  foreground pixel of its disc. Each bin is swept to a fixpoint, the eight
  bins are repeated at a level until stable, and the ascending pass over
  the levels is repeated until nothing changes. Taking the levels in
  ascending order makes boundary pixels removable before ridge pixels, so
  a medial route is kept where an edge-hugging one would otherwise
  survive.
\item \emph{Optional reduction $\mathcal R_2$} ($+R_2$ configuration only). For
  $\tau=0,1,2$ in order, a sweep on a fresh copy of the thinned skeleton
  deletes $P$ iff its 8-ring shows exactly one $0\!\to\!1$ transition,
  $t(P)=1$, with $n_4(P)<4$ (a sufficient condition for
  $(8,4)$-simplicity; end-points are not exempt), and it alone covers at
  most $\tau$ pixels of its whole disc, background pixels included:
  \begin{align}
  &n_4(P)<4,\qquad t(P)=1,\nonumber\\
  &\#\{\,q\in\mathcal{B}(P):\ c(q)=1\,\}\le\tau.
  \label{eq:r2_rule}
  \end{align}
  The result $S^{(\tau)}$ is accepted iff its reconstruction has the Euler
  number of the object,
  \begin{equation}
  \chi_{8,4}\Bigl(\Omega\cap\bigcup\nolimits_{P\in S^{(\tau)}}\mathcal{B}(P)\Bigr)
  =\chi_{8,4}(\Omega),
  \end{equation}
  with 8-connected foreground and 4-connected background; the loop stops
  at the first failing $\tau$ and returns the last accepted result, or the
  thinned skeleton if $\tau=0$ fails. SpiderSkel-DualC omits the
  $n_4(P)<4$ clause; SpiderSkel-Scaffold also never deletes a part's last
  skeleton pixel.
\end{enumerate}

\emph{The two other tails.} SpiderSkel-DualC runs steps~1, 2, 5 and~6, omits
steps~3 and~4, and in step~2 keeps one champion per $8$-connected object of
$\Omega$ rather than per $4$-connected part. SpiderSkel-Scaffold replaces
steps~1--5 by anchored erosion: the Stage-5 decision serves as bookkeeping
only, anchoring each mover at its target and every other surviving candidate
in place, plus one seed per anchor-free $4$-connected part at its first
deepest-EDT pixel in row-major order; sweeps over the foreground pixels of
$\Omega$ in ascending $d^2$, row-major within a level, delete every
non-anchor pixel that is $(8,4)$-simple in the current set, each deletion
taking effect immediately, until a sweep deletes nothing. Its thinning is
step~6 with the level condition $d^2(P)\le\lambda$ dropped.
 
\FloatBarrier

\balance

\bibliographystyle{IEEEtran}